\documentclass[english,aps,twocolumn,prd,superscriptaddress,showpacs,preprintnumbers,amsmath,amssymb]{revtex4}

\usepackage{subfigure}
\usepackage{amsmath,amssymb}
\usepackage{graphicx}
\usepackage{rotating}
\usepackage{bm}
\usepackage{dsfont}
\usepackage{color}
\usepackage[dvips]{epsfig}     
\usepackage{multirow}
\definecolor{refcol}{RGB}{0,0,205}
\usepackage{microtype}
\usepackage[colorlinks,linkcolor=refcol,citecolor=refcol,urlcolor=refcol]{hyperref}
\usepackage{breakurl}

\graphicspath{{./figures/}}

\def\di{\displaystyle}

\def\bg{\begin{eqnarray}\begin{array}{rcl}\displaystyle}
\def\eg{\end{array} &\di    &\di   \end{eqnarray}}
\def\bm#1{\begin{eqnarray}\begin{array}{#1}\di}
\def\bmo#1{\begin{eqnarray*}\begin{array}{#1}\di}
\def\bml#1#2{\begin{eqnarray}\begin{array}{#1}\label{#2}\di}
\def\bgo{\begin{eqnarray*}\begin{array}{rcl}\displaystyle}
\def\ego{\end{array} &\di    &\di \nonumber  \end{eqnarray*}}

\def\btensor#1#2{\renew\left#1\begin{array}{#2}\di}
\def\brtensor#1#2#3{\ren#3\left#1\begin{array}{#2}}
\def\botensor#1#2{\renew\left#1\begin{array}{#2}}
\def\etensor#1{\end{array}\right#1}

\def\eq#1{(\ref{#1})}
\def\Eq#1{Eq.~(\ref{#1})}
\def\Fig#1{Fig.~\ref{#1}}

\def\id{1\!\mbox{l}}

\def\s0#1#2{\mbox{\small{$ \frac{#1}{#2} $}}}
\def\0#1#2{\frac{#1}{#2}}

\def\R{{{\rm l}\!{\rm R}}}

\def\CP{{\mathcal P}}

\def\llangle{\left\langle}
\def\rrangle{\right\rangle}

\def\s{\sigma}

\def\ren#1{\renewcommand{\arraystretch}{#1}}

\def\renew{\renewcommand{\arraystretch}{1}}

\definecolor{blue}{rgb}{0,0,1}

\definecolor{green}{rgb}{0,1,0}

\definecolor{red}{rgb}{1,0,0}

\newcommand{\Tr}{\mathrm{Tr}}

\newcommand{\be}{\begin{eqnarray}}
\newcommand{\ee}{\end{eqnarray}}

\newcommand{\yb}{\bar\psi}

\newcommand{\Nf}{N_{\text{f}}}
\newcommand{\Nc}{N_{\text{c}}}
\newcommand{\lqcd}{\Lambda_{\text{QCD}}}

\usepackage{babel}
\makeatother

\begin{document}

\title{Higher order quark-mesonic scattering processes and the phase structure of QCD}

\author{Jan M.~Pawlowski}
\affiliation{Institut f\"ur Theoretische Physik, University of Heidelberg, 
Philosophenweg 16, 62910 Heidelberg, Germany}
\affiliation{ExtreMe Matter Institute EMMI, GSI, Planckstr.~1, 64291 Darmstadt, Germany.}
\author{Fabian Rennecke}
\affiliation{Institut f\"ur Theoretische Physik, University of Heidelberg, 
Philosophenweg 16, 62910 Heidelberg, Germany}
\affiliation{ExtreMe Matter Institute EMMI, GSI, Planckstr.~1, 64291 Darmstadt, Germany.}

\begin{abstract}
  We study the impact of higher order quark-meson scattering processes
  on the chiral phase structure of two-flavour QCD at finite
  temperature and quark density. Thermal, density and quantum
  fluctuations are included within a functional
  renormalisation group approach to the quark-meson model. We present
  results on the chiral phase boundary, the critical endpoint, and the
  curvature of the phase transition line at vanishing density.
\end{abstract}

\maketitle

\section{Introduction}\label{sec:Intro}

The understanding of the formation and the properties of hadronic
matter requires that of the phase structure of Quantum
Chromodynamics (QCD). For fixed density the QCD vacuum changes
drastically with decreasing temperature from a deconfined quark-gluon
plasma phase with effective chiral symmetry to a hadronic phase with
confined quarks and broken chiral symmetry. Future and running
heavy-ion experiments e.g. at CERN, FAIR, BNL and JINR aim at probing
this phase transition, especially also at high densities
\cite{BraunMunzinger:2003zd}.

The main challenge for theoretical studies of the QCD phase diagram
lies in its non-perturbative nature, and the -- related -- dynamical
change of relevant degrees of freedom. However, in the past decade
rapid progress has been made in the first principle description of QCD
at finite temperature and density, both with continuum methods, see
e.g.\
\cite{Braun:2008pi,Braun:2009gm,Pawlowski:2010ht,Herbst:2013ufa,Fischer:2011mz,%
  Fischer:2012vc,Fischer:2013eca}, and on the lattice, see e.g.\
\cite{Karsch:2001cy,Philipsen:2007rj,deForcrand:2010ys,Sexty:2013ica}.
Within the continuum approach it has been worked-out in detail how low
energy effective models are systematically embedded in first principle
QCD, see
\cite{Braun:2009gm,Pawlowski:2010ht,Kondo:2010ts,Herbst:2010rf,Herbst:2013ail%
  ,Haas:2013qwp,Herbst:2013ufa}. It is a particular strength of such
an approach that the necessary quantitative control over the matter
and glue sector can be achieved separately, followed by a systematic
combination of both sectors including their mutual back-reaction.
This puts an even bigger emphasis on the systematic improvement of the
corresponding low energy effective models of QCD. It not only furthers
our understanding of the mechanisms underlying the pyhsics phenomena
responsible for the phase structure of QCD but also is necessary for
quantitatively describing the phase structure within a first principle
continuum approach.

For small chemical potential the lightest hadronic states, the pions
and the sigma-meson, drive the chiral dynamics in the vicinity of the
phase boundary. Thus, in order to achieve quantitative control over
the matter sector of QCD, and in particular the phase stucture, one
has to accurately take into account the effects of mesonic
fluctuations. The importance of such a procedure has been already
observed in the context of higher order mesonic self-scatterings.
These have been taken into account within low energy effective models
in terms of full mesonic effective potential, for reviews see e.g.\
\cite{Berges:2000ew,Schaefer:2006sr,Braun:2011pp,vonSmekal:2012vx}. It
is also well-known that a corresponding Taylor expansion converges
well towards the results obtained with a full mesonic effective
potential \cite{Papp:1999he}. For a fully self-consistent expansion it
is important to realise that quark--anti-quark multi-meson
interactions have to be taken into account as well. Indeed, these
terms contribute directly to the computation of the effective
potential in the functional renormalisation group (FRG) approach.

Hence, in the present work, for the first time, we systematically also
include higher order quark--anti-quark multi-meson interactions within
the quark-meson (QM) model, and study their effect on the chiral phase
structure of two-flavour QCD. In the QM model this amounts to a
meson-field--dependent Yukawa coupling. The quantum, thermal, and
density fluctuations are then taken into account by means of the
FRG. This also allows us to consider the momentum-dependence of the
propagators in terms of scale-dependent wave function
renormalisations.  Such effects are particularly relevant in the
presence of massless excitations such as the pions close to a second
order phase transition. The higher quark-meson interactions are
included in a -- convergent -- Taylor expansion in the order of the
mesonic fields. In total this leads to a significant extension of the
local potential approximation of the quark meson model which has been
used extensively to study the chiral phase transition of QCD, see
e.g. \cite{Berges:2000ew,Schaefer:2006sr,Schaefer:2004en}.

We present results on the
chiral phase boundary in the $T$-$\mu$ plane, including the critical
endpoint and the curvature of the phase boundary. We also compare
different definitions of the phase boundary. This is particularly
important in the region of the phase diagram where the system
undergoes a crossover transition and the exact location of the phase
boundary is not uniquely defined. We find that the inclusion of the
higher order couplings lead to quantitatively significant changes in
the phase structure. An intriguing observation is the rapid
convergence of our results if the orders of meson-meson and
quark-meson couplings are increased. 

Our present findings are fully in the spirit of the systematic
embedding of the low energy effective models in first principle QCD. They
constitute significants steps towards quantitative precision in terms
of convergence of a self-consistent truncation for the matter sector
of QCD.

The present work is organised as follows: In section \ref{sec: model}
we briefly introduce the quark-meson model in the context of full QCD, including the higher order
quark-meson scattering processes in terms of an effective meson
potential, a field-dependent Yukawa coupling and quark and meson wave
function renormalisations. Section \ref{sec:frg} summarises the
renormalisation group approach and provides some details about the
resulting flow equations of our model. Our results are presented in
section \ref{sec: res}, where we demonstrate the convergence of our
expansion scheme and discuss the chiral phase structure at finite
temperature and quark chemical potential including the critical
endpoint and the curvature at vanishing chemical
potential. Conclusions are given in section \ref{sec: concl}. A
discussion of the expansion scheme as well as some details on the flow
equations can be found in the Appendices.

\section{Quark-Meson model}\label{sec: model}

In the present work we concentrate on two-flavour QCD by employing a
quark-meson model as a low energy effective model for QCD. As already
mentioned in the introduction, it is by now well understood how such
low-energy effective models are embedded in first-principle QCD within
functional methods. The key concept behind this embedding in full QCD
the the consistent treatment of the dynamical change of the relevant
degrees of freedom. Starting from the high temperature/large cut-off
scale quark-gluon phase, the system is dynamically driven towards the
low temperature/small cut-off scale hadronic phase, where chiral
symmetry breaking is triggered by the increasing gauge coupling. 

This transition from a description in terms of quarks and gluons to a
hadronic description is achieved by dynamical hadronisation
\cite{Gies:2001nw,Gies:2002hq,Pawlowski:2005xe,Floerchinger:2009uf}. Furthermore,
while some hadronic degrees of freedom get dynamical at the
hadronisation scale $\Lambda\! \approx\! 1\,\text{GeV}$, the quark and
gluon degrees of freedom decouple. This structure is very apparent in
the Landau gauge, where the gluon propagator is infrared gapped, the
gapping being directly related to the QCD mass gap, see
e.g. \cite{Fischer:2008uz,Maas:2011se}. Hence, the gluons can be
integrated out first, leading to an effective theory with quarks and
hadrons in a gluonic background potential, such as Polyakov-loop
enhanced low-energy models. 

This setting entails that first-principle QCD flows can be employed to
provide initial parameters and further glue input, such as background
potentials, for model calculations, thereby systematically removing
ambiguities in these models. In particular, no double counting of
degrees of freedom is present in quark-meson models in this context:
the initial parameters of the low-energy theory are fixed uniquely by
the first-principle QCD flows, and an unambiguous projection procedure for the flows of all couplings is given by a
complete orthogonal set of projection operators which is also fixed uniquely by the relation to first-principle QCD. Such a procedure
resolves for example, amongst other ambiguities, the well-known Fiertz
ambiguity, see e.g.\ \cite{Braun:2011pp}. This picture has
successfully been applied to first-principle QCD at finite temperature
and density, \cite{Braun:2009gm,Pawlowski:2010ht}, as well as to low
energy effective models
\cite{Herbst:2010rf,Herbst:2013ail,Haas:2013qwp}, where quantitative
agreement of QCD thermodynamics with lattice QCD is achieved
\cite{Herbst:2013ufa}.

Here, our focus is on the chiral dynamics of QCD at energy scales
$k\!\leq\! 700\,\text{MeV}$. In the present work, dynamical
hadronisation is not taken into account. We want to emphasise,
however, that this plays a quantitative important role for large
cutoff scales, where gluon-induced fermionic self-interactions in the
chirally symmetric phase need to be taken into account properly. Since
we start the RG-flow of our model at a scale
$\Lambda\!=\!700\,\text{MeV}$, where composite degrees of freedom have
already formed, dynamical hadronisation would only lead to very minor
quantitative corrections. We have explicitly checked this statement
for simpler QM models. We have also neglected the gluonic background
which leads to a quantitative change of the phase boundary. This will
be discussed elsewhere.

\subsection{Low energy effective action}\label{sec:lowenergy} 
For an effective description of the low energy matter sector of QCD at
not too high densities, a model based solely on quarks and the lightest meson is a
good approximation. The ultraviolet cut-off scale $\Lambda$ of such a
description, as already mentioned above, relates to the scale where the pure glue sector of QCD
decouples and the fluctuations of the lightest mesons and constituent
quarks dominate the dynamics. Here we consider $\Nf=2$ degenerate
quark flavours with pseudo-scalar pions $\vec{\pi}$ and the scalar
sigma meson as the dominant mesonic degrees of freedom for not too
large chemical potential at $\Lambda\approx 1$ GeV. At this scale the
low energy effective action $\Gamma_\Lambda$ is approximated by that
of the quark-meson model. 

As described in the previous section, for momentum
scales $k\lesssim \Lambda$, we take into account the scale-dependent
dressing of the propagators and higher mesonic-- as well as
quark-meson--interactions. The corresponding scale-dependent effective
action reads
\begin{eqnarray}\nonumber 
  \Gamma_k &= &\int_x
\Bigl\{iZ_{\psi,k}(\rho)
  \yb(\gamma_\mu \partial_\mu+\gamma_0\mu)\psi +\frac{1}{2}
Z_{\phi,k}(\rho)(\partial_\mu\phi)^2 \Bigr.
\\ &&\Bigl.
 +V_k(\rho)-c
  \sigma+ h_k(\rho)\yb(\gamma_5\vec{\tau} \vec{\pi}+i\sigma)\psi \Bigr\}\,, 
\label{eq:model}\end{eqnarray}
with the meson fields in the $O(4)$ representation, $\phi = (\vec{\pi},\sigma)$, and 
\begin{equation}\label{eq:rho} 
\rho
= \frac{1}{2}\phi^2=\frac{1}{2}(\vec{\pi}^{2}+\sigma^2)\,. 
\end{equation} 
In \eq{eq:model} we used the abbreviation $\int_x = \int_0^{1/T} d x_0
\int d^3 x$. The quark fields $\psi$ are two flavour Dirac-spinors and
$\mu$ is the quark chemical potential. For $\Nf=2$ the chiral symmetry
$SU(\Nf)_L \otimes SU(\Nf)_R$ is isomorphic to $SO(4)$, hence the
$O(4)$-symmetry of the scalar effective potential
$V_{k}(\rho)$. Quarks and mesons are coupled via a meson
field-dependent Yukawa coupling $h_{k}(\rho)$. $\vec{\tau}$ are the
Pauli matrices.

This model captures spontaneous chiral symmetry breaking $SU(\Nf)_L
\otimes SU(\Nf)_R \rightarrow SU(\Nf)_V$. The expectation value of the
sigma meson serves as order parameter for the chiral phase transition
and the three pions are Goldstone bosons of the spontaneous breaking
of the axial $SU(2)_A$. In the presence of explicit symmetry breaking,
introduced by the linear breaking term $-c\sigma$, the pions are not
massless but rather pseudo-Goldstone bosons with finite mass and the
chiral second order transition turns into a crossover.

The inverse quark and meson propagators are dressed with wave function
renormalisations $Z_{\psi,k}(\rho)$ and $Z_{\phi,k}(\rho)$. Note that
at finite temperature there are in general two different wave function
renormalisations quarks and mesons, one perpendicular, $Z=Z^\perp$,
and one parallel to the heat bath, $Z^{\|}$. In the present work we
only compute the perpendicular one and identify
$Z^\|=Z^\perp$. Moreover, we expect a weak dependence of the $Z$'s on
the meson field $\rho$ and hence we drop all terms proportional to
$\partial_\rho Z(\rho)$. This approximation is discussed in section
\ref{sec: etaflow}. The explicit breaking of $O(4)$-symmetry in the meson-sector 
of our model through the linear term $-c \sigma$ is related to a finite current 
quark mass $m_q^c$ via the relation
\begin{align}\label{eq: cqm}
m_q^c = \frac{h_\Lambda}{\lambda_{1,\Lambda}} c\, ,
\end{align}
where $\lambda_{1,\Lambda}$ is the squared meson mass in the UV, see (\ref{vansatz}).

\subsection{Higher order mesonic scattering}\label{sec: messcat}

The present approximation includes field-dependent wave function
renormalisations $Z_k(\rho)$ for quarks and mesons, a full effective
potential $V_k(\rho)$, and a field-dependent Yukawa-coupling
$h_k(\rho)$. We implement higher order mesonic scattering processes via a systematic expansion in $n$-point functions
$\Gamma^{(n)}$ of the effective action \eq{eq:model}.

We first discuss the wave function renormalisations.  The
$\rho$-dependence of the mesonic wave function renormalisation
contains momentum-dependent meson self-interactions while that of the
quarks contains momentum-dependent scattering of a quark--anti-quark
pair with mesons. Note that both processes vanish at vanishing
momenta. The wave function renormalisations can be expanded about a
temperature and chemical potential dependent and potentially
scale-dependent expansion point $\kappa_k(T,\mu)$, to wit
\begin{equation}\label{eq: zexp}
Z_k(\rho)=\sum_{n=0}^{N_Z}\frac{Z_{n,k}}{n!}\left(\rho-\kappa_k(T,\mu)
\right)^n\,.
\end{equation}
However, we expect a rather mild dependence of the wave function
renormalisation on the meson field, leading to
\begin{equation}\label{eq:locallyfield}
\partial_\rho Z_k(\rho)\approx 0\,. 
\end{equation} 
The quantitative reliability of this hypothesis is tested in
Appendix~\ref{app: exp}, see in particular 
\Fig{fig: epsdep}. \Eq{eq:locallyfield} implies that locally (about a given 
expansion point $\kappa_k$) we can use 
\begin{align}\label{eq: zev}
Z_k =Z_{0,k}\,.  
\end{align} 
Still, for the computation of observables the wave function
renormalisation has to be determined at the expectation value $\rho_0$
of the mesonic field which does not necessarily agree with the
expansion point $\kappa_{k=0}$. It is here where the field dependence
of the $Z$'s play a r$\hat {\rm o}$le. 

Meson self-interactions are contained in the effective potential
$V_{k}(\rho)$. As for the wave function renormalisations we expand
the effective potential in powers of $\rho$ about an 
expansion point $\kappa_k(T,\mu)$, to wit
\begin{align}\label{vansatz}
  V_{k}(\rho) = \sum_{n=1}^{N_V}\frac{\lambda_{n,k}}{n!}
  \bigl(\rho-\kappa_k\bigr)^n\,. 
\end{align}
In \eq{vansatz} we have dropped all $T,\mu$-dependence for the sake of
brevity. \Eq{vansatz} captures a chiral crossover and a second order
transition for $N_V \geq 2$.  A first order transition requires at
least $N_V \geq 3$. The effect of higher order mesonic
self-interactions on the matter sector of QCD can be systematically
studied by increasing the order of the expansion $N_V$.

It is convenient to rewrite the effective action in terms of the
renormalised fields 
\begin{equation}\label{eq:renfields} 
\bar\phi= Z_{\phi,k}^{1/2}\, \phi\,,\qquad {\rm and} \qquad \bar\rho = 
Z_{\phi,k}\,\rho\,,  
\end{equation} 
where the wave function renormalisations are locally constant in the
present approximation. For the effective potential $\bar
V(\bar\rho)=V(\rho)$ this implies
\begin{equation}\label{eq:vbar}
 \bar  V_{k}(\bar \rho) = \sum_{n=1}^{N_V}\frac{\bar \lambda_{n,k}}{n!}
  \bigl(\bar \rho-\bar\kappa_k\bigr)^n.
\end{equation}
with 
\begin{equation}\label{eq:barlambda} 
  \bar{\lambda}_{n,k}=\0{\lambda_{n,k} }{Z_{\phi,k}^{n}} \,,
  \qquad {\rm and}\qquad \bar\kappa_k= 
  Z_{\phi,k}\,\kappa_k\,.
\end{equation}
The $\bar\rho$-derivatives of the effective potential,
$\partial_{\bar\rho}^n \bar V$, and that of the effective action allow
for a direct physical interpretation as they are RG-invariant. The
linear term in \eq{eq:model} reads in the new fields
\begin{equation}\label{eq:barc} 
c\sigma =\bar c_k \bar \sigma\,,\qquad {\rm with}
\qquad \bar c_k=\0{c}{Z_{\phi,k}^{1/2}}\,. 
\end{equation}
Quark--multi-meson interactions are taken into account in
(\ref{eq:model}) by the coupling of two quarks and a meson with a
$\rho$-dependent Yukawa coupling $h_{k}(\rho)$. Analogous to the
effective potential, we also expand the Yukawa coupling in a
$O(4)$-symmetric manner in powers of $(\rho-\kappa_k)$,
\begin{align}\label{hansatz}
h_{k}(\rho) = \sum_{n=0}^{N_h}\frac{h_{n,k}}{n!}\bigl(\rho-\kappa_k\bigr)^n\,.
\end{align}
$N_h=0$ amounts to the standard Yukawa interaction which couples a quark--anti-quark pair and a meson. By increasing $N_h$ the interaction between a quark--anti-quark pair and $(2 N_h+1)$ mesons can be taken into account. The renormalised analogue of \eq{hansatz}
reads
\begin{equation}\label{eq:barh}
\bar h_{k}(\bar \rho) = \0{h_{k}(\rho)}{Z_{\psi,k} Z_{\phi,k}^{1/2}} 
 = \sum_{n=0}^{N_h}\frac{\bar h_{n,k}}{n!}\bigl(\bar \rho-\bar\kappa_k\bigr)^n\,, 
\end{equation}
with RG-invariant expansion coefficients 
\begin{equation}\label{eq:barhn}
\bar{h}_{n,k}=\0{h_{n,k} }{Z_{\psi,k} Z_{\phi,k}^{(2n+1)/2 }} \,.
\end{equation}
The convergence of these expansions implies that the higher order
couplings get increasingly irrelevant with increasing order of the meson field, see section \ref{sec: conv}. A
more detailed analysis of the present expansion scheme is deferred to
Appendix~\ref{app: exp}. Here we only note that we choose a scale
independent expansion point $\kappa$.

\section{Fluctuations}\label{sec:frg}

In the present work we include quantum, thermal and density
fluctuations with the functional renormalisation group (FRG). In
addition to its application to QCD, see
\cite{Litim:1998nf,Pawlowski:2005xe,Gies:2006wv,Pawlowski:2010ht,Rosten:2010vm} and
corresponding low enegry effective models
\cite{Berges:2000ew,Schaefer:2006sr,Braun:2011pp,vonSmekal:2012vx},
the FRG has been used successfully in a variety of physical problems
ranging from ultracold atoms and condensed matter physics
\cite{Delamotte:2007pf,Metzner:2011cw,Boettcher:2012cm,Braun:2011pp}
to quantum gravity
\cite{Niedermaier:2006wt,Codello:2008vh,Litim:2011cp,%
  Reuter:2012id}.
\begin{figure}[t]
\includegraphics[height=8.2ex]{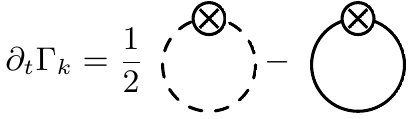}
\caption{Diagrammatic representation of the flow equation for the
  matter sector of QCD. The dashed line represents the full meson
  propagator, the solid line the full quark propagator and the crossed
  circle depicts the regulator insertion.}
\label{fig: fleq}
\end{figure} 
The idea is to start with the effective action $\Gamma_\Lambda$ given
in \eq{eq:model} at the initial scale $k=\Lambda$ and to successively
include quantum fluctuations by integrating out momentum shells down
to an infrared-cutoff scale $k$. By lowering $k$ we resolve the
macroscopic properties of the system and eventually arrive at the full
quantum effective action $\Gamma$ at $k=0$. This evolution of
$\Gamma_k$ is governed by the Wetterich equation
\cite{Wetterich:1992yh}. For the quark-meson model it reads,
\begin{align}\label{fleq}
\begin{split}
  \partial_t\Gamma_k[\phi,\psi,\yb]=&\frac{1}{2}\text{Tr}\!\left[\left( 
\0{1}{\Gamma_k^{(2)}[\phi,\psi,\yb]+R_k}\right)_{\phi\phi}
\partial_t
R_k^\phi\right] \\&-\text{Tr}\!\left[\left( \0{1}{\Gamma_k^{(2)}[\phi,\psi,\yb]+
  R_k}\right)_{\psi\yb}\partial_t R_k^\psi\right] \,,
\end{split}
\end{align}
where $\partial_t$ is the total derivative with respect to the RG-time
$t=\ln( k/\Lambda)$. The traces sum over the corresponding discrete
and continuous indices including momenta and species of
fields. $\Gamma_k^{(2)}[\phi,\psi,\yb]$ is the matrix of second
functional derivatives of $\Gamma_k$ with respect to the
fields. The indices $\phi\phi$ and $\psi\yb$ indicate the components in field space. In this notation the regulator $R_k$ is also a matrix in field space, where $R_k^\phi$ and $R_k^\psi$ are the entries corresponding to the meson and quark regulators
respectively. The flow equation has a simple diagrammatic
representation, see \Fig{fig: fleq}.

The specific regulators used in the present work are
three-dimensional Litim regulators and are specified in \eq{eq:regs}
in Appendix~\ref{app:ThresholdFunc}. They are proportional to the wave
function renormalisations $Z_\phi,Z_\psi$ and hence are RG-adapted,
\cite{Pawlowski:2001df,Pawlowski:2005xe,Gies:2002af}. As a consequence
of this choice the $Z$'s completely drop out of the flows and only the
anomalous dimensions survive. Note that this only holds true within
the locally constant approximation for the wave function
renormalisation. It also implies that our regulators depend on the
expansion point. Note also that fully field-dependent wave function
renormalisations $Z(\rho)$ cannot be introduced to the regulators
without modifying the flow equation, see \cite{Pawlowski:2005xe}. Such
a flow will be considered elsewhere. The regulators suppress modes
with momenta $p^2\lesssim k^2$ and thus implement the successive
inclusion of fluctuations on ever-lower energy scales down to the
scale $k$.

On the one hand such an approximation to the full QCD-flow is only
satisfactory in the low energy regime of QCD at scales $k\lesssim 1$
GeV and k should be chosen as small as possible. On the other hand the
initial cut-off scale $\Lambda$ has to be far bigger than any other
physical scale under investigation, i.e. $T$, $\mu$ and the physical
masses. In the present work we shall adopt $\Lambda=700\,
\text{MeV}$. Note that $\Lambda$ receives a physical meaning in this
context since it is directly related to the scale where hadrons form.

It is left to project the flow equation (\ref{fleq}) for the effective
action on the scale- and field-dependent parameters of the effective
action defined in \eq{eq:model}: 

\subsection{Effective potential}

The flow equation of the effective potential $V_k(\rho)-c\sigma$ is
obtained by evaluating \eq{fleq} for constant meson fields,
$\phi(x)\rightarrow\phi$, and vanishing quark fields. For these field
configurations the effective action reduces to
$\Gamma_k=\text{Vol}_4^{-1} \left( \bar V_k(\bar \rho)-\bar c_k\bar
  \sigma\right)$, see \eq{eq:model}.  Note that the explicit symmetry
breaking term is linear in the meson field, and hence is nothing but a
source term. The right hand side of the flow equation \eq{fleq} only
involves second derivatives w.r.t.\ the fields. Thus, the explicit
symmetry breaking term does not appear on the right hand side of the
flow equation, which only depends on symmetric terms. Moreover the
flow equation \eq{fleq} is derived with cut-off--independent source
terms, which implies $\partial_t c=0$. With \eq{eq:barc} this leads to
\begin{align}\label{eq: cflow}
\partial_t \bar{c}_k=\frac{1}{2}\eta_{\phi,k}\,\bar{c}_k\,,
\end{align}
with the (perpendicular) meson anomalous dimension
\begin{align}
\eta_{\phi,k} =-\frac{\partial_t Z_{\phi,k}}{Z_{\phi,k}}\,.
\end{align}
This has the remarkable consequence, that in terms of fluctuations the
theory is effectively evaluated in the chiral limit. Hence, this also applies to the
effective potential $V_k$. The explicit
$O(4)$ symmetry breaking introduced via the linear term $c\,\sigma$
simply entails that the vacuum expectation value of the fields is shifted relative to that in the chiral
limit. In other words, the physical observables that can be derived
from $V_k$ and its derivatives are evaluated away from the minimum of
$V_k$. Note also in this context that we could choose any
$k$-dependence for $c_k$,  only the value at $k=0$ is fixed
by the physical quark masses.  The flow for $\bar V_k(\bar
\rho)=V_k(\rho)$ reads
\begin{align}\label{vflow}
\begin{split}
\hspace{-.4cm}\left.   \partial_t\right|_{\rho} \bar V_k(\bar{\rho}) =& 
\0{k^4}{4 \pi^2} \left\{ \left[(\Nf^2-1)\, l_0^{(B,4)}(\bar{m}_{\pi,k}^2,
\eta_{\phi,k};T)\right.\right.\\ 
&\left.+l_0^{(B,4)}(\bar{m}_{\sigma,k}^2,\eta_{\phi,k};T)
\right]\\
&\left.- 4N_c N_f\,
    l_0^{(F,4)}(\bar{m}_{\psi,k}^2,\eta_{\psi,k};T,\mu)
  \right\}\,,
\end{split}
\end{align}
with the (perpendicular)
quark anomalous dimension,
\begin{align}\label{eq: etadef}
 \eta_{\psi,k}=-\frac{\partial_t Z_{\psi,k}}{Z_{\psi,k}}\,.
\end{align}
\Eq{vflow} is nothing but the flow equation $\partial_t V(\rho)$.  The
threshold functions $l_0^{B/F,4}$ are defined in
Appendix~\ref{app:ThresholdFunc}, and depend on the field-dependent
dimensionless renormalised masses
\begin{align}\label{eq:mmass}
\bar m_{\pi,k}^2 = \0{ \partial_{\bar \rho} \bar V_k}{k^2} \,,
\qquad\qquad  \bar m_{\sigma,k}^2 = \0{\partial_{\bar \rho} \bar V_k 
+2 \rho\,\partial^2_{\bar \rho} \bar V_k}{ k^2}\,,
\end{align}
and 
\begin{align}\label{eq:qmass}
\bar m_{\psi,k}^2=\0{2 \bar h_k(\bar \rho)^2 \bar \rho}{k^2}\,.
\end{align}
The first and second lines in (\ref{vflow}) are the pion and the sigma
meson contributions respectively. The third line in (\ref{vflow}) is
the quark contribution, where $2 N_c N_f$ is the number of internal
quark degrees of freedom. The additional factor $-2$ is generic for
fermionic loops. 

The flow of the renormalised couplings $\bar\lambda_n$
is derived from \eq{eq:vbar} as ($n\geq 1$),
\begin{align}\label{eq:lflow}
\begin{split}
&\left. \partial^n_{\bar \rho}\left.   \partial_t\right|_{\rho} \bar V_k(\bar{\rho})
\right|_{\bar\rho=\bar\kappa_k}\\
&\qquad=\left( \partial_t -n\,\eta_{\phi,k}\right)\bar{\lambda}_{n,k} -\bar{\lambda}_{n+1,k}\,\left(
\partial_t+\eta_{\phi,k} \right)\bar\kappa_k \,.
\end{split}
\end{align}
where we have used that \eq{eq:renfields}  implies 
\begin{equation}
\partial_t\bar\rho = -\eta_{\phi,k} \bar\rho\,. 
\end{equation}
We have computed the scale derivative at fixed $\rho$ as we want to connect
\eq{eq:lflow} to \eq{vflow}: the left hand side of \eq{eq:lflow} is
the $n$th derivative w.r.t.\ $\bar\rho$ of the flow equation
\eq{vflow}. The relative sign for the $\eta_{\phi,k}$-terms in the second line 
of \eq{eq:lflow} reflects the factor $1/Z_{\phi,k}$ in the renormalised couplings 
in comparison to the factor $Z_{\phi,k}$ in the expansion point $\bar\kappa$. 

The equations \eq{eq:lflow} with \eq{vflow} provide a tower of coupled
differential equations for higher order mesonic correlators and
therefore include meson-meson scattering up to order $2 N_V$ into our
model. We now use the freedom of choosing an expansion point
$\bar\kappa$ in order to improve the convergence property of the
Taylor expansion. To that end we note that an expansion about fixed
$\bar\rho=\bar\kappa$ with $\partial_t \bar\kappa=0$ keeps a term
proportional to $\bar\lambda_{n+1}$ on the right hand side of
\eq{eq:lflow}. Such a linear dependence of the flow of $\bar\lambda_n$
potentially destabilises the expansion as the $\bar\lambda_{n+1}$ grow
rapidly with $n$ even though their relevance for the potential decreases 
rapidly. This is indeed the case as we have checked within an
explicit numerical computation. Note that this argument also applies to an expansion about the scale dependent minimum of the effective potential $\bar\rho_{0,k}$.

In turn, the linear $\bar\lambda_{n+1}$-contribution vanishes precisely for 
\begin{align}\label{eq: kappaflow}
  \partial_t \bar\kappa_k =- \eta_{\phi,k} \bar\kappa_k\,,\qquad \to
  \qquad \partial_t\kappa = 0\,.
\end{align}
that is a scale-independent bare expansion point. We emphasise that it
is only the expansion about a fixed bare field value that removes the
destabilising back-coupling of the higher order couplings. Therefore, it is
the preferred expansion point in terms of convergence of the
expansion. An explicit numerical check indeed reveals the rapid
convergence of such an expansion, see Section~\ref{sec:
  conv}. However, we have also checked that both expansions
convergence to the same results.

We close this Section with a discussion of possible order
parameters. For a large region of the phase diagram the chiral
transition is a cross-over. This only allows for the definition of a
pseudo-critical temperature which is not unique. All possible
definitions of pseudo order parameters have the property that they
provide order parameters in the chiral limit where the cross-over
turns into a second order phase transition. Here we discuss several 
order parameters. The variance of the pseudo-critical temperatures provide 
a measure for the width of the cross-over. 

A simple order parameter of the chiral transition (in the chiral
limit) is given by the vacuum expectation value $\bar \sigma_{0,k}$ at
vanishing cut-off. It also determines the pion decay constant,
$f_\pi=\bar \sigma_{0,k=0}$. The expectation value $\bar \sigma_{0,k}$
is obtained from
\begin{align}\label{eq: min}
\Bigl.\partial_{\bar \rho}\left[ \bar V_k(\bar\rho)-
\bar c_k \bar \sigma \right]\Bigr|_{\bar \rho=\bar \rho_{0,k}}=0,
\end{align}
where $\bar\rho_{0,k}=\frac{1}{2} \bar\sigma_{0,k}^2$ is the quadratic
order parameter. Physical observables such as the pion decay constant
and the masses are then defined at vanishing cut-off scale $k=0$ and
$\bar\rho = \bar\rho_{0,\text{IR}}$.

The position of the peak of the chiral susceptibility is an
alternative definition of the phase transition temperature. The chiral
susceptibility measures the strength of chiral fluctuations. Hence it
is, independent of its use for constructing an order parameter, an
interesting observable. It is defined as the response of
the chiral condensate $\llangle \bar\psi \psi \rrangle$ to variations
of the current quark mass $m_q^c$,
\begin{align}\label{eq: chisu}
\chi_{\bar q q} = \frac{\partial\! \llangle \bar\psi \psi \rrangle}{
  \partial \bar m_q^c}.
\end{align}
Within our model the scale dependent chiral condensate is given by
\cite{Ellwanger:1994wy, Berges:1997eu}
\begin{align}\label{eq: chic}
  \llangle \bar\psi \psi \rrangle_k = -\frac{1}{
    h_{\Lambda}}\left( \lambda_{1,\Lambda} \sigma_{0,k} -
    c\right).
\end{align}
The current quark mass is given by (\ref{eq: cqm}). Combining \eq{eq:
  chic} and \eq{eq: cqm} yields the following relation:
\begin{align}\label{eq: sich}
  \frac{\partial\sigma_{0,\text{IR}}}{\partial c} =
  -\frac{h_\Lambda^2}{ \lambda_{1,\Lambda}^2}\chi_{\bar q
    q}+\frac{1}{\lambda_{1,\Lambda}}.
\end{align}
Note that for \eq{eq: chic} to hold with high accuracy, we need to
require that the expansion point $\kappa$ is very close to the
physical point $\rho_{0,k}$. This is indeed the case for our choice of
the expansion point, see \eq{eq: epd}.

We can rewrite \eq{eq: sich} by virtue of the implicit function
theorem since the relation between $\sigma_{0,k}$ and $c$ is
implicitly given by \eq{eq: min}. This yields:
\begin{align}
  \frac{\partial\sigma_{0,k}}{\partial c} =\Bigl( V'_k(\rho_{0,k})+
  2\rho_{0,k}V''_k(\rho_{0,k}) \Bigr)^{-1}=\frac{1}{m_{\sigma,k}^2}.
\end{align}
Thus, in practice we can compute the sigma meson mass and readily
extract the chiral susceptibility for given initial parameters
$\lambda_{1,\Lambda},\,h_\Lambda$.

\subsection{Yukawa coupling}\label{sec: hflow}

\begin{figure}[t]
\includegraphics[height=8.2ex]{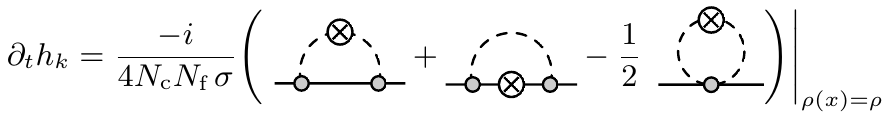}
\caption{Diagrammatic representation of the flow of the Yukawa
  coupling. The grey circles depict the full vertices.}
\label{fig: hflow}
\end{figure}

The scalar part of the Yukawa-term has been introduced in
\eq{eq:model} as a $\phi$-dependent fermionic mass term with mass
$h(\rho) \sigma$. This definition also entails that in leading order
in $\rho$ the sigma field has been introduced as a field for the
composite operator $\bar\psi\psi$.  Accordingly we evaluate the flow
of the fermionic two-point function at the minimal fermionic momentum 
$p_{\rm low}$ and constant mesonic fields, leading to
\begin{align}\label{eq: hproj}
\begin{split}
  \partial_t h_k(\rho)=&-\frac{1}{\sigma}\frac{i}{4\Nc\Nf}\\
  &\times\text{Re}\left[\lim_{p\rightarrow
      p_\text{low}}\text{Tr}\left.\left(
        \frac{\delta^2\partial_t\Gamma_k}{\delta\psi(-p)\delta\yb(p)}
       \right)\right|_{\rho(x)=\rho}\right],
\end{split}
\end{align}
where the trace  in \eq{eq: hproj} sums over all internal indices. Note that \eq{eq: hproj} is 
well-defined even in the limit $\sigma\to 0$. The diagrammatic representation 
of this equation is depicted in \Fig{fig: hflow}. 

In \eq{eq: hproj} we have set the external spatial momenta to zero and
the external Matsubara frequencies to their lowest mode,
$p_\text{low}=(\pi T,\vec{0}\,)$ for quarks. Implicitly we also use
$p_\text{low}=(0,\vec{0}\,)$ for mesons as we evaluate the Yukawa
coupling for constant mesonic fields. For finite quark chemical
potential this procedure yields a manifestly complex valued flow on
the right hand side of \eq{eq: hproj}. This simply reflects the
dependence of the two-point function on $p_0- i\mu$ and hence the
momentum-dependence of the Yukawa coupling $h$: Evaluated at constant
mesonic fields the Yukawa coupling is a function of $\rho$, $(p_0-
i\mu)^2$, $\vec p^2$ and $\mu$ with real expansion coefficients:
$h(p_0- i\mu)^*=h(p_0+ i\mu)$. Hence, any projection procedure has to
reflect the property that 
\begin{equation}\label{eq:real}
h( p_0- i\mu) + h(- p_0- i\mu)\in \R\,,
\end{equation}
where we have also used that the Yukawa coupling $h$ is a function of
$(p_0- i\mu)^2$. \Eq{eq:real} has to hold in any self-consistent
approximation scheme. In the present derivative expansion the Yukawa
coupling is evaluated at a fixed frequency. This means that the Yukawa
coupling in the derivative expansion has to be chosen real, $h( p_0- i\mu) = h(- p_0- i\mu)$. Within the
flow this can be achieved via an appropriate choice of the expansion
point. This singles out vanishing frequency $p_0=0$, where
\eq{eq:real} holds trivially. 

More generally one can project the flow of the Yukawa coupling on its
real part. The former projection procedure at vanishing frequency has
been used in the literature, for a detailed discussion and motivation
of this approach see \cite{Braun:2008pi}. However, the latter procedure keeps
the Matsubara mass-gap of the fermions, which also is potentially
relevant for capturing the quantitative physics close to the Fermi
surface of the quarks at higher density. Hence, in the present work we
project on the real part of the flow in \eq{eq: hproj} for the
computation of the Yukawa coupling. We have checked numerically 
that both procedures agree quantitatively for small chemical potential.

The projection (\ref{eq: hproj}) using the fermionic two-point
function is directly related to the more customary projection where an
additional derivative with respect to the pion fields is applied. One
finds
\begin{align*}
  -\frac{i}{\sigma} \text{Tr}\left(\partial_t
    \Gamma_k^{(\psi\yb)}\right) = \left.\text{Tr}\left(\gamma_5
      \vec{\tau}\,\partial_{\vec{\pi}}\partial_t
      \Gamma_k^{(\psi\yb)}\right)\right|_{\vec{\pi}=0}\,.
\end{align*}
Note that a projection using an additional derivative with respect to
the sigma field would contaminate the flow with additional
contributions from the derivative of the Yukawa coupling.

With (\ref{eq: hproj}) we find for the flow of the renormalised
Yukawa coupling:
\begin{align}\label{hflow}
\begin{split}
  &\left.\partial_t\right|_\rho \bar{h}_k(\bar{\rho})=\\
&\qquad\Bigl(
  \frac{1}{2}\eta_{\phi,k}+ \eta_{\psi,k}
  \Bigr)\bar{h}_k(\bar{\rho})\\
  &\qquad+4 v_3 \bar{h}_k^3(\bar\rho) \left[
    L_{(1,1)}^{(4)}\left(\bar{m}_{\psi,k}^2,\bar{m}_{\sigma,k}^2,
      \eta_{\psi,k},\eta_{\phi,k};T,\mu\right)\right.\\
  &\qquad\left.-(\Nf^2-1)\, L_{(1,1)}^{(4)}\left(\bar{m}_{\psi,k}^2,\bar{m}_{\pi,k}^2,
      \eta_{\psi,k},\eta_{\phi,k};T,\mu\right)\right]\\
  &\qquad+16 v_3\bar{h}_k(\bar{\rho}) \bar{h}'_k(\bar{\rho})\bar{\rho}
  \Bigl[\bar{h}_k(\bar{\rho})+\bar\rho\bar{h}'_k(\bar{\rho})\Bigr]\\
  &\qquad\times
  L_{(1,1)}^{(4)}\left(\bar{m}_{\psi,k}^2,\bar{m}_{\sigma,k}^2,
    \eta_{\psi,k},\eta_{\phi,k};T,\mu\right)\\
  &\qquad-2v_3k^2 \Bigl[\left(3\bar{h}'_k(\bar{\rho})+2 \bar{\rho}
    \bar{h}''_k(\bar{\rho})\right) l_1^{(B,4)}(\bar{m}_{\sigma,k}^2,
  \eta_{\phi,k};T)\Bigr.\\
  &\qquad\Bigl.+ 3\bar{h}'_k(\bar{\rho})\,
  l_1^{(B,4)}(\bar{m}_{\pi,k}^2,\eta_{\phi,k};T)\Bigr]\,.
\end{split}
\end{align}
The function $L_{(1,1)}^{(4)}$ is defined in
Appendix~\ref{app:ThresholdFunc}. We note that the terms proportional
to $\bar{h}_k^3$ in eq. (\ref{hflow}) are the triangle-diagram
contributions to the flow of a field-independent Yukawa coupling, see
e.g.~\cite{Braun:2008pi}. For the renormalised couplings \eq{eq:barhn}
in (\ref{eq:barh}) we find analogously to \eq{eq:lflow}
\begin{align}\label{eq: hcflow}
\begin{split}
&\left.\partial_{\bar\rho}^n
\left.\partial_t\right|_\rho \bar{h}_k(\bar{\rho})\right|_{\bar\rho=\bar\kappa_k}\\
&\qquad= (\partial_t-n\, \eta_{\phi,k}) \bar{h}_{n,k}-\bar h_{n+1,k} (\partial_t+\eta_{\phi,k}) 
\bar\kappa_k\,,
\end{split}
\end{align}
where $\partial_t \bar{h}_k(\bar{\rho})$ is given by
eq. (\ref{hflow}). Hence the flows of $\bar h_{n,k}$ show the same
decoupling properties within the expansion scheme already discussed 
below \eq{eq:lflow}. 

\subsection{Wave function renormalisations}\label{sec: etaflow}

As discussed at the end of Section~\ref{sec:lowenergy},
at finite temperature the wave function renormalisations perpendicular
and parallel to the heat bath differ from each other, $Z_k^\perp\!
\neq\!  Z_k^\parallel$. For scales above the chiral symmetry breaking
scale, $k>k_{\chi\text{SB}}$, we have $T/k<1$ which implies that
thermal fluctuations are negligible and thus
$Z_{k>k_{\chi\text{SB}}}^\perp \approx
Z_{k>k_{\chi\text{SB}}}^\parallel$. In the infrared we have $k\!\ll\! T$. In this regime dimensional reduction
occurs and we approach the three-dimensional limit. There the finite
temperature RG flow is only driven by the lowest Matsubara modes. The
lowest Matsubara mode for bosons is zero and therefore $Z_{\phi,k\ll
  T}^\parallel$ drops out. For fermions the lowest Matsubara mode is
proportional to $T$ and thus the fermions with dynamically generated
mass effectively decouple from the flow in the infrared. Therefore we
choose the approximation $Z_k\!=\!Z_k^\perp\!  =\! Z_k^\parallel$,
which is approximately valid for large scales and hardly affects the
RG flow in the infrared. Hence it should be a good approximation for
calculating the chiral phase boundary. 
%
\begin{figure}[t]
\includegraphics[height=8.2ex]{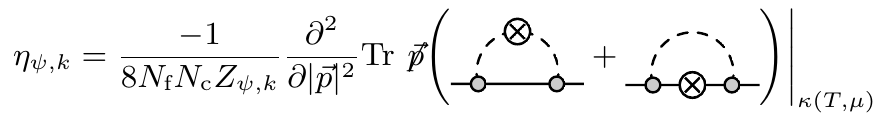}
\caption{Diagrammatic representation of the quark anomalous dimension.}
\label{fig: etapsi}
\end{figure} 
%

The flow of \eq{eq: zev} consistent with the expansion scheme about
$\rho=\kappa$ has to involve an evaluation of the two-point function
at the expansion point.  As the momentum dependence is covered in a
coarse-grained form via the $k$-dependence of the $Z_k$'s, we also use
the derivative expansion about the lowest momentum and frequency and
arrive at
\begin{eqnarray}\label{eq:ano}
\eta_{\psi,k}&=& -\frac{1}{8 \Nf \Nc Z_{\psi,k}}\\ & &\hspace{-.5cm} \!\times
\text{Re}\left[\lim_{p\rightarrow p_\text{low}}\frac{\partial^2}{\partial |
\vec{p}|^2}\text{Tr}\left.\left(\vec{\gamma}\vec{p} \,\frac{\delta^2
    \partial_t \Gamma_k}{\delta\psi(-p)\delta\yb(p)} \right)
\right|_{\rho=\kappa}\right]\, ,
\nonumber 
\end{eqnarray} 
for the anomalous dimension $\eta_{\psi,k}$ defined in (\ref{eq:
  etadef}). We note that, analogous to the computation of the Yukawa
coupling, the projection onto external momentum $p_\text{low}$ also
renders the flow on the right hand side complex valued. Similarly to
the Yukawa coupling, the anomalous dimension is a function of the
complex variable $(p_0 - i \mu)$, and projecting onto the real part, see
\eq{eq:ano}, keeps all properties and symmetries intact. This leads to
\begin{align}\label{etapsi}
\nonumber\eta_{\psi,k} = &\frac{2 v_3}{3}\left(4-\eta_{\phi,k}\right)\\
&\nonumber\times\Bigl[(\Nf^2-1)\, \bar{h}_{k}(\bar\kappa_k)^2\mathcal{FB}_{(1,2)}\left(\bar{m}_{\psi,k}^2,
\bar{m}_{\pi,k}^2;T,\mu\right) \Bigr.\\
&+\Bigl.\Bigl( \bar{h}_{k}(\bar\kappa_k)+2 \bar\kappa_k\,\bar{h}_{k}'(\bar\kappa_k) 
\Bigr)^{\!2}\Bigr.\\
&\nonumber\times\, \Bigl. \mathcal{FB}_{(1,2)}\left(\bar{m}_{\psi,k}^2,
\bar{m}_{\sigma,k}^2;T,\mu\right)  \Bigr].
\end{align}
The function $\mathcal{FB}_{(1,2)}^{(4)}$ is defined in
Appendix~\ref{app:ThresholdFunc}. In the case of one quark flavour and
for $\bar{h}_{k}'=0$ this equation agrees with that found in
\cite{Braun:2009si}. The diagrammatic representation of equation
(\ref{etapsi}) is shown in \Fig{fig: etapsi}.

%
\begin{figure}[t]
\includegraphics[height=8.2ex]{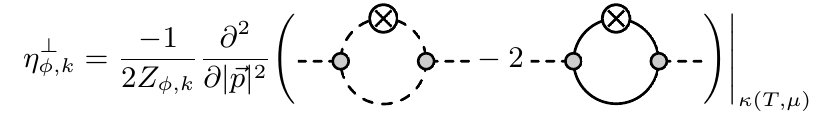}
\caption{Diagrammatic representation of the meson anomalous
  dimension.}
\label{fig: etaphi}
\end{figure}
%

The anomalous dimension of the mesons can be either extracted by
taking derivatives w.r.t.\ $\sigma$ or $\pi$. Here, a similar argument as for the flow of the Yukawa
coupling (Section~\ref{sec: hflow}) applies and we choose the following projection:
\begin{align}\label{etaphi proj}
  \eta_{\phi,k}=-\frac{1}{2 Z_{\phi,k}}\lim_{p\rightarrow
    p_\text{low}}\frac{\partial^2}{\partial |\vec{p}|^2}
  \text{Tr}\left.\left(\frac{\delta^2\partial_t \Gamma_k}{
        \delta\pi_i(-p)\delta\pi_i(p)}
    \right)\right|_{\rho=\kappa(T,\mu)},
\end{align}
where the choice of $i=1,2,3$ does not matter since the pions always
have $O(3)$ symmetry in this representation. This leads to 
\begin{align}\label{etaphi}
\begin{split}
  \eta_{\phi,k}=&\frac{8 v_3}{3}\Bigl\{2 k^{-2} \bar\kappa_k\,
  (\bar{V}_{k}''(\bar\kappa_k))^2\, \mathcal{BB}_{(2,2)}\left(\bar{m}_{\pi,k}^2,
    \bar{m}_{\sigma,k}^2;T,\mu\right) \Bigr.\\
  &+\Bigl. \Nf\Nc \bar{h}_{k}(\bar\kappa_k)^2\left[\left(2
      \eta_{\psi,k}-3\right)\mathcal{F}_{(2)}(\bar{m}_{\psi,k}^2;T,
    \mu)\right.\Bigr.\\
  &-\Bigl.\left.4\left(\eta_{\psi,k}-2\right)
    \mathcal{F}_{(3)}(\bar{m}_{\psi,k}^2;T,\mu) \right] \Bigr\}.
\end{split}
\end{align}
The functions $\mathcal{BB}_{(2,2)}^{(4)}$ and
$\mathcal{F}_{(n)}^{(4)}$ are defined in
Appendix~\ref{app:ThresholdFunc}. This equation also agrees with
\cite{Braun:2009si} for one quark flavour. We note that the results
shown here are obtained using the optimised regulator shape functions
(\ref{litR}). The diagrammatic representation of (\ref{etaphi}) is
shown in \Fig{fig: etaphi}.

We want to emphasize again that the wave function renormalisations are
defined as the zeroth order of an expansion about $\bar\kappa_k$. From
\eq{etapsi} and \eq{etaphi} it is clear that we can define the
wave function renormalisations at any expansion point. Defining them
at $\bar\kappa_k$ is the consistent way to define the renormalized
couplings related to the Yukawa coupling and the effective potential
since these couplings are defined at $\bar\kappa_k$ as well.  For the
definition of the physical parameters, e.g. the masses, however, we
make use of the option to freely choose the expansion point and
evaluate the wave function renormalisations at the minimum
$\bar\rho_{0,k}$, see appendix \ref{app: exp}.

\begin{figure}[t]
\begin{center}
  \includegraphics[width=.95\columnwidth]{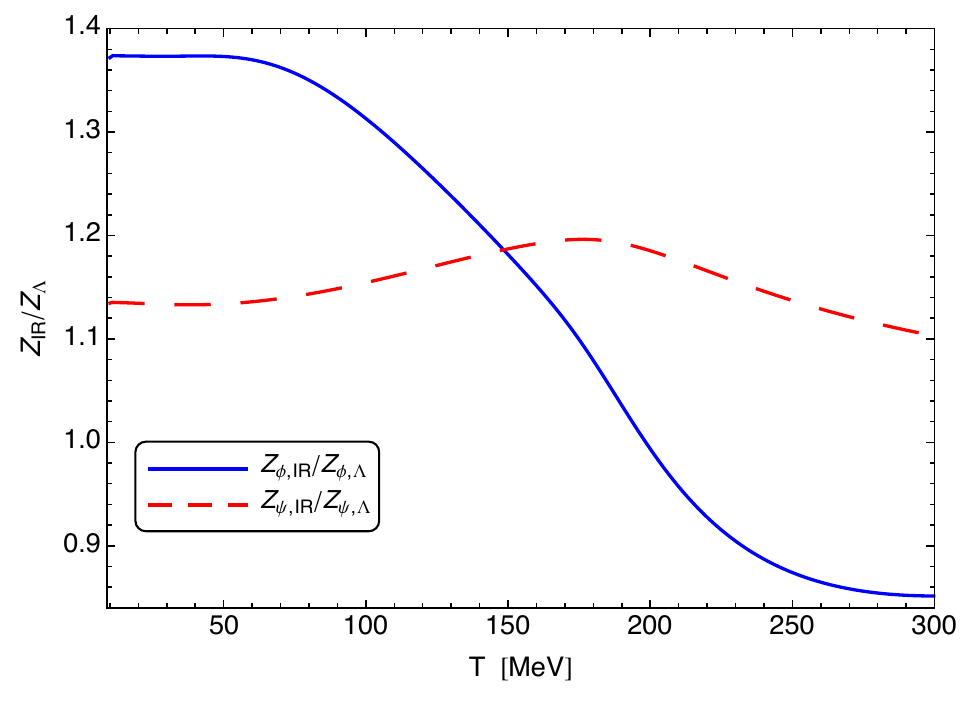}
  \caption{Temperature dependence of the wave function
    renormalisations in the IR, normalised to their values in the
    UV.}\label{fig: Zs}
\end{center}
\end{figure}

The wave function renormalisations as a function of temperature are
shown in \Fig{fig: Zs}. At about the critical temperature
$Z_{\psi,\text{IR}}$ exhibits a peak and $Z_{\phi,\text{IR}}$ shows a
tiny kink. This kink gets more pronounced and turns into a dip for
smaller pion masses and ends up as a non-analyticity in the chiral
limit \cite{Berges:1997eu}. An interesting observation is that the
meson wave function renormalisation falls below its initial value in
the UV for temperatures above $200$ MeV. This feature is independent
of the choice of the UV value and shows that mesonic degrees of
freedom become less important for larger temperatures in the crossover
region and vanish in the symmetric phase. The inclusion of a dynamical
meson wave function renormalisation therefore leads to a consistent
picture of the QCD matter sector in the sense that mesons are only
present in the phase with broken chiral symmetry, while they vanish
(or rather turn into auxiliary fields) in the symmetric phase where
quarks and gluons are the relevant degrees of freedom.

\subsection{Convexity for $\rho<\rho_0$ \& behaviour for large
  fields}\label{subs:conlarge}

The flows are initiated at a UV cut-off scale $k=\Lambda$. There, the
initial effective action $\Gamma_\Lambda$ resembles the classical
Yukawa theory with a $\phi^4$-potential and a constant Yukawa
coupling. This entails that all flows decay in the limit where
$\rho/k^2\to\infty$, and we conclude
\begin{align}\label{hrlarge}
\begin{split}
  \partial_t \bar{h}_k(\rho/k^2\rightarrow\infty)=0\, \quad {\rm
    and} \quad \partial_t \bar V_k(\rho/k^2\rightarrow\infty)=0\,. 
\end{split}
\end{align}
Hence, neither the Yukawa coupling nor the effective potential are
changed during the flow for large enough fields. 

For the convexity discussion it is sufficient to restrict ourselves to
the chiral limit. As already mentioned before the explicit symmetry
breaking does not influence the dynamics of the fluctuations which are
solely responsible for the convexity properties. It is well-known that
the effective potential $V_{k=0}$ is convex by construction. It has
been shown that the flow equation is indeed convexity-restoring in the
limit $k\to 0$ and keeps this property in the local potential
approximation, see \cite{Litim:2006nn}. This implies in particular,
that the the curvature of the effective potential vanishes for field
values smaller than their vacuum expectation value, $\partial_\phi^2
V_{k=0}(\rho\!<\!\rho_0)=0$. For
non-vanishing cut-off scale, $k>0$, only the combination $V_k(\rho) +
\rho\,R_k^\phi(0) $ is convex. This property follows directly from its
definition as the Legendre transformation of the logarithm of the
generating functional $\ln Z_k[J]$, evaluated on constant fields and
divided by the space-time volume, for a detailed discussion see e.g.\
\cite{Pawlowski:2005xe}. This entails in particular
\begin{equation}\label{eq:sing} 
V'_k(\rho) + R_k^\phi(0)\geq 0 \,,\quad V'_k(\rho) +2 \rho
V''(\rho) +R_k^\phi(0)\geq 0\,, 
\end{equation}
for the inverse propagator pion and $\sigma$-meson propagator respectively at
vanishing momentum. In the chirally broken phase with
$\rho<\rho_{0,k}$ we have
\begin{equation}\label{eq:negative} 
V'_k(\rho<\rho_{0,k})<0 \quad {\rm and} \quad V'_{k=0}(\rho<\rho_0)=0\,, 
\end{equation} 
with $\rho_0$ is the vaccuum expextation value $\rho_{0,k}$ at
vanishing cut-off scale, $k=0$. Note that \eq{eq:negative} also
implies $V''_{k=0}(\rho<\rho_0)=0$. Of course, \eq{eq:negative}
entails that the potential $V_0$ is flat (vanishing curvature) for
fields $\rho$ smaller than the vaccuum expectation value. Note that
for negative $V'$ the inverse pion propagator in \eq{eq:sing} vanishes
for $V'=R_k^\phi(0)$, and the flow potentially diverges for
$R_k^\phi(0)\to 0$. However, this divergence is not reached as the
increasing flow increases $V'$, see \cite{Litim:2006nn} for a
discussion of a scalar theory. In Appendix~\ref{app:con} this
discussion is extended to the present Yukawa theory.

\begin{figure*}[t]
\begin{center}
\subfigure{\includegraphics[width=0.48\textwidth]{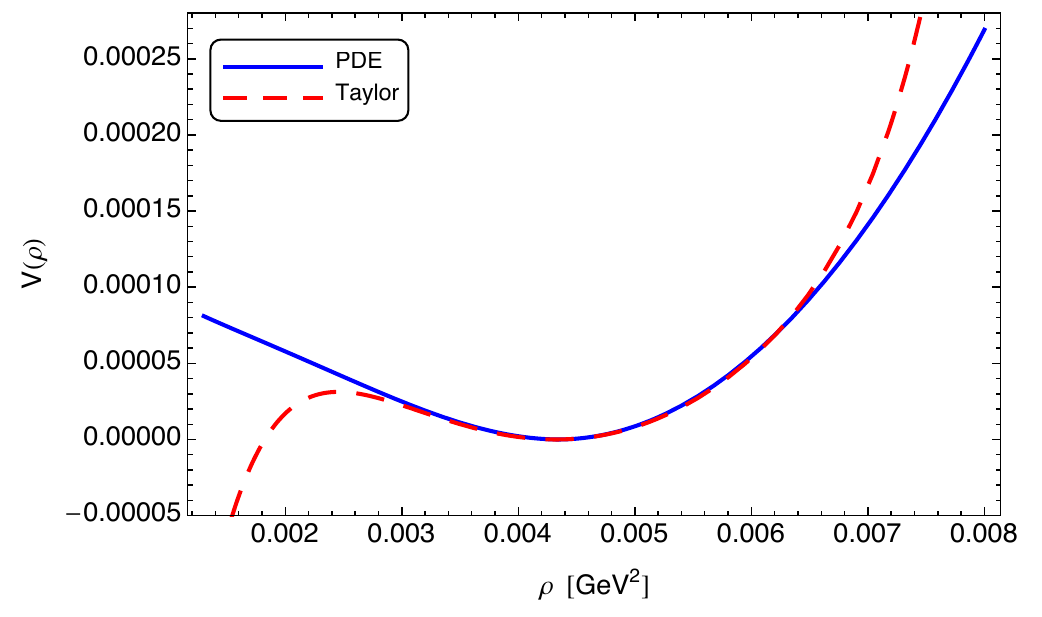}}
\hspace{0.02\textwidth}
\subfigure{\includegraphics[width=0.48\textwidth]{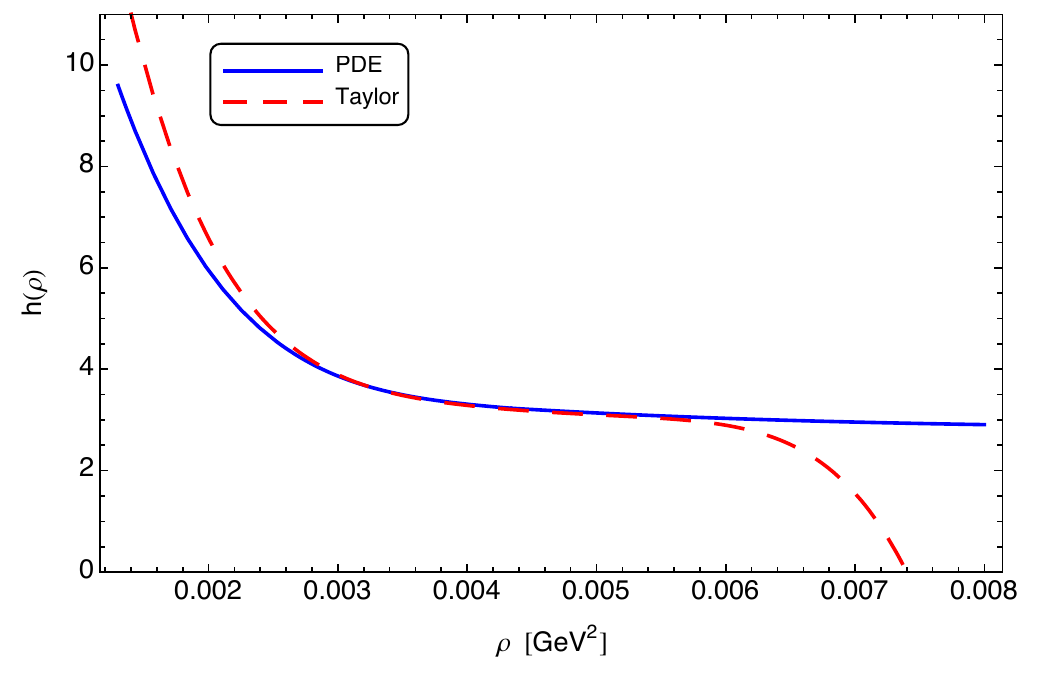}}
\caption{Effective potential (left) and Yukawa coupling (right) at
  cut-off scale $k\!=\!0.1\,\text{MeV}$ within the Taylor expansion
  and the full solution of the coupled partial differential flow
  equations for $\eta_{\psi,k},\eta_{\phi,k}$, $V_k(\rho)$ and $h_k(\rho)$ at $T\!=\!10\,\text{MeV}$
  and $\mu\!=\!0$. For the Taylor expansion we used $N_V=7$ and
  $N_h=5$. Note that the results include the explicit $O(4)$ symmetry
  breaking via $-c\sigma$.}\label{fig:PDEvsTaylor}
\end{center}
\end{figure*}

The relation between the negative curvature $V'$ and the
convexity-restoration in the flow also implies that only the mesonic
fluctuations drive the flows for $\rho<\rho_0$ and $k\to 0$, and the
two sectors effectively decouple. This facilitates the access to the 
infrared flow of the fermion propagator in this region studied in detail in 
Appendix~\ref{app:con}, which can indeed be derived analytically. 
We arrive in particular at 
\begin{equation}\label{eq:ferm2p=0}
  \bar m_\psi^2(\rho\leq \rho_0)=\sqrt{2 \rho_0}  
\bar h (\rho_0,0) \0{\rho_0}{\rho}\,, 
\end{equation}
see \eq{eq:ferm2p=0app} in Appendix~\ref{app:con}. In this appendix a
momentum-dependent Yukawa coupling $h(\rho,p)$ has been introduced. 
In \eq{eq:ferm2p=0} it is evaluated at vanishing momentum, $h(\rho,0)$. 
This leads to the
inequality
\begin{equation}\label{eq:massgap} 
\bar m_\psi^2(\rho) \geq \bar m^2_{\psi,\rm gap}\,,  \quad {\rm with}\quad  
\bar m^2_{\psi,\rm gap}= \sqrt{2 \rho_0}   \bar h (\rho_0,0)\,, 
\end{equation} 
for $k=0$, see \eq{eq:massgapapp} in Appendix~\ref{app:con}, where we
have also used the fact that for $\rho>\rho_0$ the mass function
grows. As a consequence of \eq{eq:massgap} the fermion propagation is
gapped with at least the constitutent quark mass $\bar m^2_{\psi,\rm
  gap}$ for all fields. In other words, in the chirally broken phase
no mesonic background can turn the fermionic dispersion into a
massless one. Note however, that $\rho<\rho_0$ is no physical choice
in the first place.

We finally remark that the same line of argument can also be applied
to full QCD and also holds there. There however, the fermionic mass
tends towards the current quark masses for large meson fields $\rho$
and the minimum in \eq{eq:massgap} has to be restricted to
$\rho\lesssim \rho_0$.  In the present model a linearly rising mass
was built-in and strictly speaking one should not evaluate the model
for $\rho/\Lambda^2 \gg 1$ in the first place.  The full discussion of
QCD is postponed to future work.

\section{Set-up and Results}\label{sec: res}

\subsection{Initial conditions in the UV}\label{subs: ini}

It is left to specify the initial conditions
for the relevant parameters $\lambda_{1,k},\lambda_{2,k}, h_{0,k}$ and
$c_k$ at the UV-scale $\Lambda$ for the system of coupled flow
equations \eq{eq: cflow}, \eq{eq:lflow}, \eq{eq: kappaflow}, \eq{eq:
  hcflow}, \eq{etapsi} and \eq{etaphi}. As we have mentioned in
section \ref{sec: model}, the effective UV-cutoff scale $\Lambda$ has
a direct physical meaning in our setting. It is the scale where the
dominant part of the gluonic degrees of freedom has been integrated
out and hadronic degrees of freedom, especially the light mesons,
form. There is a certain freedom in the choice of this scale as long
as it is well above $\lqcd$ and not too large so that fluctuations in
the gauge sector dominate the dynamics. This requirements bring forth
$\Lambda \in [0.6,1]\, \text{GeV}$ as an approximate window for the
choice of the UV-cutoff.

As discussed in Section~\ref{sec:frg} we have chosen $\Lambda=700\,
\text{MeV}$.  The relevant parameters of our model are fixed such that
a specific set of vacuum low-energy observables is reproduced in the
in the IR. These observables are the pion decay constant $f_\pi$, the
renormalized sigma and pion masses $\bar m_\sigma,\, \bar m_\pi$ and
the constituent quark mass $\bar m_q$ of the degenerate up and down
quarks. The explicit symmetry breaking is related to the pion decay
constant and the pion mass via $\bar c_k = \bar m_\pi^2 f_\pi$ and the
relations of our parameters to the quark and mesons masses are shown
in eq. (\ref{eq:qmass}) and (\ref{eq:mmass}). At a large scale
$\Lambda$ the model is quasi-classical and hence we choose
\begin{align}
\begin{split}
\bar V_\Lambda(\bar\rho) &= \frac{\bar\lambda}{2}(\bar\rho-\bar\nu)^2\\
\bar h_\Lambda(\bar\rho) &= \bar h =\text{const.}
\end{split}
\end{align}
The underlying assumption is that at $\Lambda$ the dynamics are
controlled by the leading order processes, i.e. the four-meson and the
quark-antiquark-meson scattering. The higher order couplings are
generated at lower scales $k<\Lambda$. We indeed found that the higher
order operators, i.e. $\bar\lambda_{n,k}$ with $n\geq 3$ and $\bar
h_{m,k}$ with $m\geq 1$ are generated at $k\lesssim 400\, \text{MeV}$,
which is well below our choice for the UV-cutoff. Since the higher
order operators are not present at our initial scale, the scale where
they are generated is a prediction of our model.

In order to reproduce the vacuum IR-observables listed above we used
the following initial values: $\bar\lambda = 71.6,\, \bar\nu = 0,\,
\bar h = 3.6,\, \bar c_\Lambda = 2.1\cdot 10^{-3}\,
\text{GeV}^3$. These initial values result in the following values for
the vacuum IR-observables, 
\begin{align}\label{eq:vac}
\begin{split}
f_\pi &= 93.0\, \text{MeV}\\
\bar m_\pi &= 138.7\, \text{MeV}\\
\bar m_\sigma &= 538.2\, \text{MeV}\\
\bar m_q &= 298.3\, \text{MeV},
\end{split}
\end{align}
which are in good agreement with their values provided by the Particle
Data Group \cite{Beringer:1900zz}. The initial values of the
parameters for the present computations are chosen such that they
reproduce the vacuum physics displayed in \eq{eq:vac} for $T,\mu=0$
and vanishing cut-off for the fully field-dependent effective
potential $\bar V_k(\bar\rho)$ and Yukawa coupling $\bar
h_k(\bar\rho)$, including running wave function renormalisations
$Z_{\phi,k}$ and $Z_{\psi,k}$. With the convergence pattern discussed
in the next section it is sufficient to use $N_V=7$ and $N_h=5$ in the
expansions \eq{vansatz} and \eq{hansatz}, and fix the initial
parameters for these values.

The bare expansion point $\kappa(T,\mu)$ is chosen to be
scale-independent. We take it close to the IR-minimum of the effective
potential for every temperature and chemical potential:
\begin{align}\label{eq: epd}
\bar\kappa_\text{IR}(T,\mu)=(1+\epsilon)\,\bar\rho_{0,\text{IR}}(T,\mu),
\end{align}
where $\epsilon>0$ gives a small offset that guarantees that
$\bar\kappa_k(T,\mu)$ is always slightly larger than the minimum of
the effective potential $\bar\rho_{0,k}$ and does not lie in the flat
region of the convex effective potential $V_{k=0}$. The details are
deferred to Appendix~\ref{app: exp}. It can be read-off from \Fig{fig:
  epsdep} that a quantitative agreement of the physics results is
obtained for expansion points in the range 
\begin{align}\label{eq: expansion}
0\leq \epsilon \lesssim 1\,. 
\end{align}
This self-consistency check within the present expansion scheme is
impressively sustained by the comparison with the full solution of the 
system of partial different equations for $V_k(\rho)$, $h_k(\rho)$, $\eta_{\phi,k}(\bar\kappa_k)$ and $\eta_{\psi,k}(\bar\kappa_k)$, see \Fig{fig:PDEvsTaylor}. The region where the results from the Taylor expansion and the full solution of the partial differential equation agree give an estimate for the radius of convergence of the Taylor expansion. This is in agreement with the study of the robustness of the expansion in Appendix~\ref{app: exp} and in particular with \Fig{fig: epsdep}.


\subsection{Effect of higher order mesonic interactions on the chiral
  order parameter}\label{sec: conv}

\begin{figure*}[t]
\begin{center}
\subfigure{\includegraphics[width=0.48\textwidth]{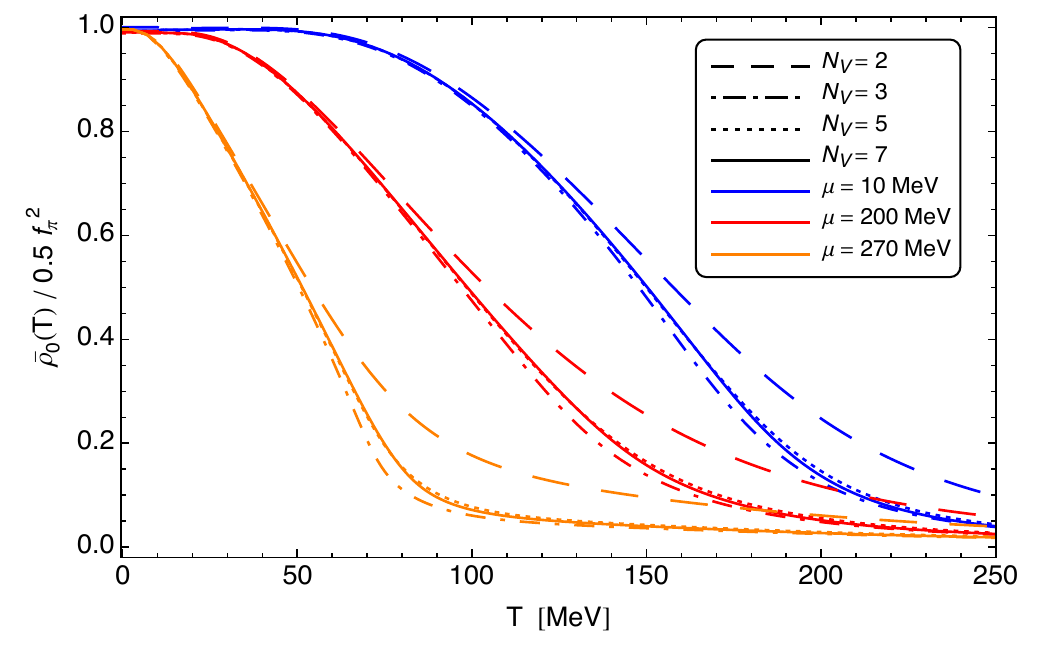}}
\hspace{0.02\textwidth}
\subfigure{\includegraphics[width=0.48\textwidth]{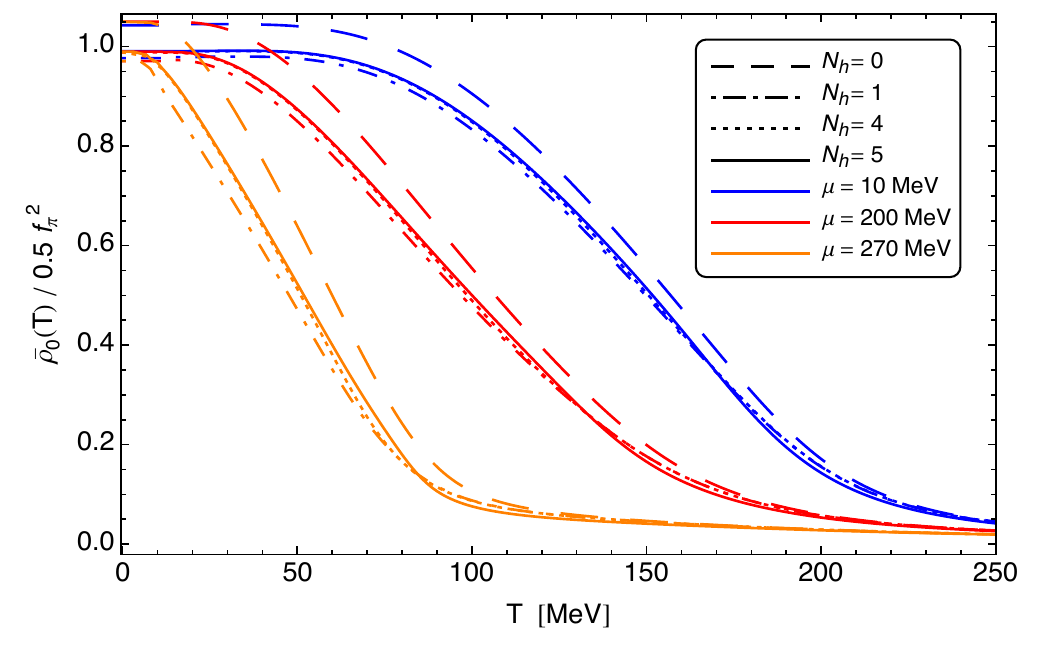}}
\caption{The normalised chiral order parameter as a function of temperature for
  different orders in the expansion of the effective potential (left)
  and the Yukawa coupling (right).}\label{fig: conv}
\end{center}
\end{figure*}

The effect of higher order mesonic operators is studied within a
Taylor expansion of the effective potential and the field dependent
Yukawa coupling, see section \ref{sec: messcat}. This is done
by comparing the results of the chiral order parameter for different
orders $N_V$ and $N_h$ in the expansions (\ref{vansatz}) and
(\ref{hansatz}) of the effective potential and the field dependent
Yukawa coupling. In \Fig{fig: conv} we show the effect
of increasing $N_V$ and $N_h$ on the chiral order parameter
$\bar\rho_{0,\text{IR}}$ as function of temperature for three
different chemical potentials.

First of all, we clearly see spontaneous chiral symmetry breaking.
Owing to the explicit symmetry breaking, the chiral condensate is very
small, but nonzero at large temperatures. By lowering the temperature,
the fluctuations of the light current quarks drive the system
continuously towards the broken phase. As the value of the chiral
condensate increases, the quarks receive more and more constituent
mass while the pions get lighter until quarks and mesons decouple at
low temperatures where the flow stops and the system ends up in the
stable phase with broken chiral symmetry. The quark and meson masses
as a function of temperature at $\mu=0$ are shown in \Fig{fig:
  masses}. Note that the decreasing slope of the meson masses at temperatures $T\gtrsim 250\,\text{MeV}$ is a result of thermal fluctuations which become of the order of the UV-cutoff $\Lambda$ in this region. This is discussed in detail in \cite{Herbst:2013ufa}.

With increasing quark chemical potential, quark fluctuations are
enhanced and the crossover gets steeper while the transition moves
towards smaller temperatures as a result of the higher quark
density.

Note that since the transition is a cross-over the actual value of the
critical temperature $T_c$ depends on the the definition of the
crossover. In this case it is only sensible to speak about a
transition region. Therefore the full temperature and chemical
potential dependence of the observables used to define the critical
temperature plays a more important role than the specific critical
values.

%
\begin{figure}[b]
\begin{center}
  \includegraphics[width=.9\columnwidth]{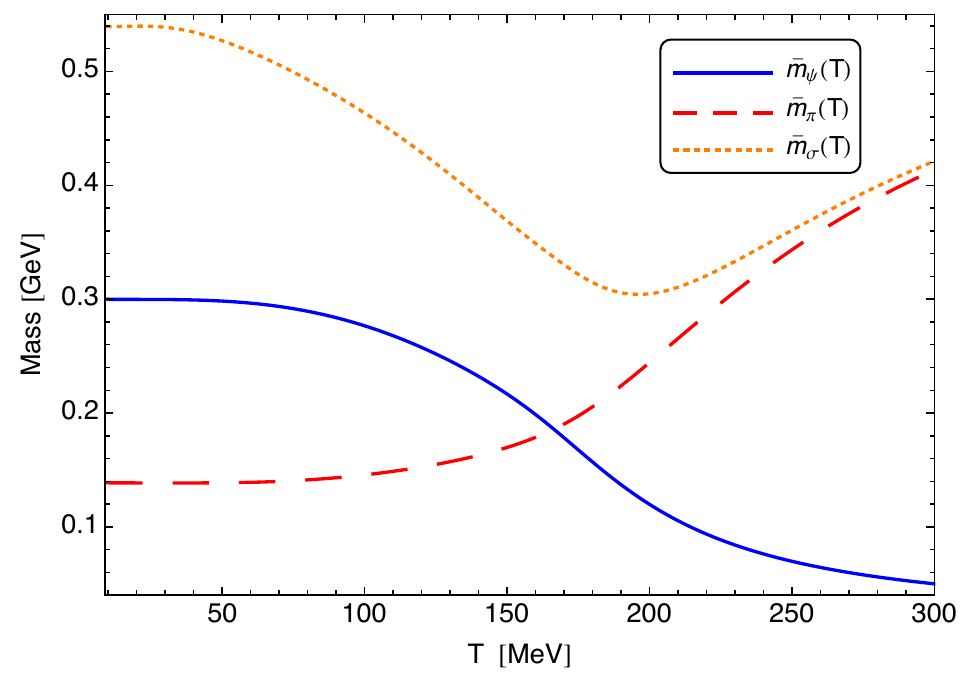}
  \caption{Quark, pion and sigma mass as a function of Temperature for
    vanishing chemical potential.}\label{fig: masses}
\end{center}
\end{figure}
%

The left panel in \Fig{fig: conv} shows the chiral order parameter in
the IR normalised to the pion decay constant for different orders
$N_V=2,\,3,\,5,\,7$ of the expansion of the effective potential for
$\mu=10\,\text{MeV},\,200\,\text{MeV},\,270\,\text{MeV}$ and a fixed
order $N_h=2$ in the expansion of the Yukawa coupling. Note that we
chose $N_h$ such that $N_h \leq N_V$ for numerical stability. While
$\rho_0(T)$ is hardly affected by different $N_V$ in the broken phase
at small temperatures, we see a large difference between the $\phi^4$
and the $\phi^{14}$ expansion in the lower region of the crossover
transition. This effect gets more pronounced for larger chemical
potentials. There is a very good agreement between the order parameter
for $N_V=5$ and $N_V=7$ which implies that the expansion of the
effective potential at order $N_V=5$ has converged to a precision of
the critical temperature below $1$ MeV. We explicitly checked that
larger orders in the expansion do not spoil this observation.

The effect of the expansion of the field dependent Yukawa coupling on
the chiral condensate is shown in the right panel of \Fig{fig:
  conv}. The difference between the usual running Yukawa coupling
$N_h=0$ and the expansion of order $N_h=5$ is at about $8\%$ which
results in a difference of $8\!-\!10$ MeV in the critical
temperature. The expansion of order $N_h=4$ seems to be converged to a
precision of the critical temperature below $1$ MeV. We observed that
larger chemical potential slows down the convergence of the Yukawa
coupling. This behaviour is expected since a larger chemical potential
effectively increases quark fluctuations and thus the systems is more
sensitive to the details of the quark-meson interactions.

We see that the particular meson-meson and quark-meson interactions we
have chosen here have a large quantitative effect on the chiral order
parameter.  Moreover we nicely see that these higher order operators a
become increasingly irrelevant with increasing order of the meson
fields and that our expansion converges rapidly, especially for not
too large chemical potential. This implies in particular that we have
the full effective potential as well as the full field-dependent
Yukawa coupling in this region. To demonstrate this, we solved the
coupled partial differential flow equations of $V_k(\rho)$, $h_k(\rho)$, $\eta_{\phi,k}(\bar\kappa_k)$ and $\eta_{\psi,k}(\bar\kappa_k)$ and compared the result
to the one obtained with the expansion employed in this work, see
\Fig{fig:PDEvsTaylor}.

Note that, as expected, the couplings
with negative mass dimension run into a Gaussian fixed point in the IR 
but certainly play a role at intermediate scales. Furthermore, it is
obvious that a low-order expansion is not sufficient in order to
obtain a high degree of quantitative precision.

\subsection{Phase diagram}

%
\begin{figure}[t]
\begin{center}
  \includegraphics[width=1.01\columnwidth]{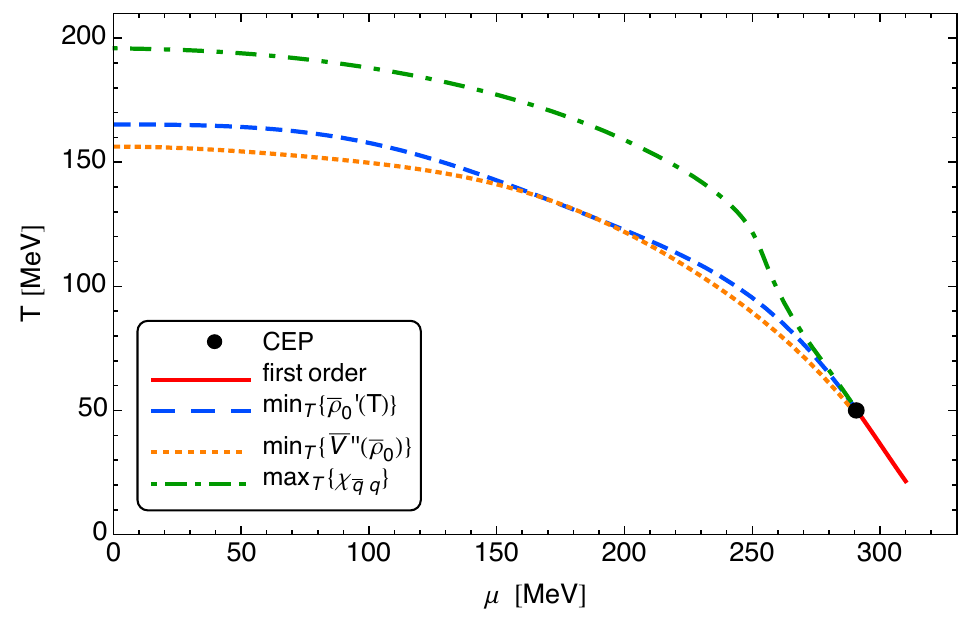}
  \caption{The phase diagram of the chiral transition including the
    different definitions of the crossover transition line we used. We
    only show the first order transition up to $\mu=310\, \text{MeV}$
    since our expansion is not fully converged for larger $\mu$, see
    Appendix~\ref{app: exp}.}\label{fig: pb}
\end{center}
\end{figure}

\subsubsection{Phase structure}\label{sec: phasestruc}

For the computation of the phase diagram, we expand the effective
potential up to order $N_V=7$ and the Yukawa coupling up to
$N_h=5$. According to the previous section, these orders guarantee
that both expansions converged to a precision below 1 MeV, at least in
the crossover region. The resulting phase diagram in the
$(T,\mu)$-plane is shown in \Fig{fig: pb}. The crossover transition
temperature is not uniquely defined and therefore depends on the
observable used to define it. Basically any observable that exhibits a
non-differentiable behaviour at the critical temperature in the chiral
limit, where the transition is of second order, can be used to define
the crossover transition temperature. Here we use the following three
definitions:
\begin{itemize}
\item[(i)] The inflection point of the chiral order parameter as a
  function of temperature,
 \begin{align}
\min_T\left\{\frac{\partial \bar\rho_{0,\text{IR}}}{\partial T}\right\}.
 \end{align}
\item[(ii)] The minimum of the quartic meson coupling at the physical point,
\begin{align}
\min_T\left\{\left.\frac{\partial^2 \bar{V}_\text{IR}(\bar\rho)}{
      \partial \bar\rho^2}\right|_{\bar\rho=\bar\rho_{0,k}}\right\}.
 \end{align}
\item[(iii)] The maximum of the chiral susceptibility (\ref{eq: chisu}),
\begin{align}
\max_T\left\{\chi_{\bar{q}q}\right\}.
 \end{align}
\end{itemize}  
The definition (i) is commonly used in RG-studies of the phase
diagram, while susceptibilities as in (iii) are typically used in lattice gauge
theory. The exact location and in particular the curvature of the
phase boundary obviously depend on the definition of the
crossover. Note, however, that all the definitions above exactly agree
in the chiral limit.

We observe a large difference of about 40 MeV in the critical
temperature at small and intermediate chemical potential between
definitions (i) and (iii), while (i) and (ii) give similar phase
boundaries. These differences are related to the fact that we have a
very broad crossover in this region and the notion of a phase
transition line is certainly not well defined there. At large chemical
potential close to the critical point the crossover lines merge and
give a uniquely defined phase boundary. This behaviour is expected
since the crossover gets steeper towards the critical point and the
first-order transition is uniquely defined. We find the critical
endpoint at $(T_\text{CEP},\mu_\text{CEP}) =(50,291)\,
\text{MeV}$. The critical endpoint here is at substantially smaller
temperatures as in mean-field studies, see
e.g. \cite{Schaefer:2007pw}. This nicely demonstrates the effect of
fluctuations on the phase boundary. The critical temperatures at
vanishing chemical potential for the different definitions of the
crossover transition are show in table \ref{tab: tc}.
\begin{table}[ht]
\begin{tabular}{|c| c|}
  \hline
  boundary def. & $T_c\,[\text{MeV}]$  \\
  \hline\hline
  (i) & 166\\
  \hline
 (ii) & 156\\
  \hline
 (iii) & 196\\
  \hline
\end{tabular}
\caption{Critical temperatures at vanishing quark chemical 
potential for the different definitions of the crossover 
phase boundary we used in this work (see text).}
\label{tab: tc}
\end{table}

A further definition of a cross-over temperature in the literature is given by the
temperature where the value of the normalised order parameter is half
of that at vanishing temperature, $\rho_{0,\text{IR}}(T,\mu)/\rho_{0,\text{IR}}(0,0)=0.5$. 
Here we only note that the critical temperature at $\mu\!=\!0$
is $T_c\! =\! 152$ MeV and the transition line is systematically below
the lines shown in \Fig{fig: pb}. This behavior is in contrast
to studies of the quark-meson model in the local potential
approximation, where this phase boundary is always slightly above the
boundary defined by (i), see e.g. \cite{Herbst:2013ail}.

%
\begin{figure}[t]
\begin{center}
  \includegraphics[width=.95\columnwidth]{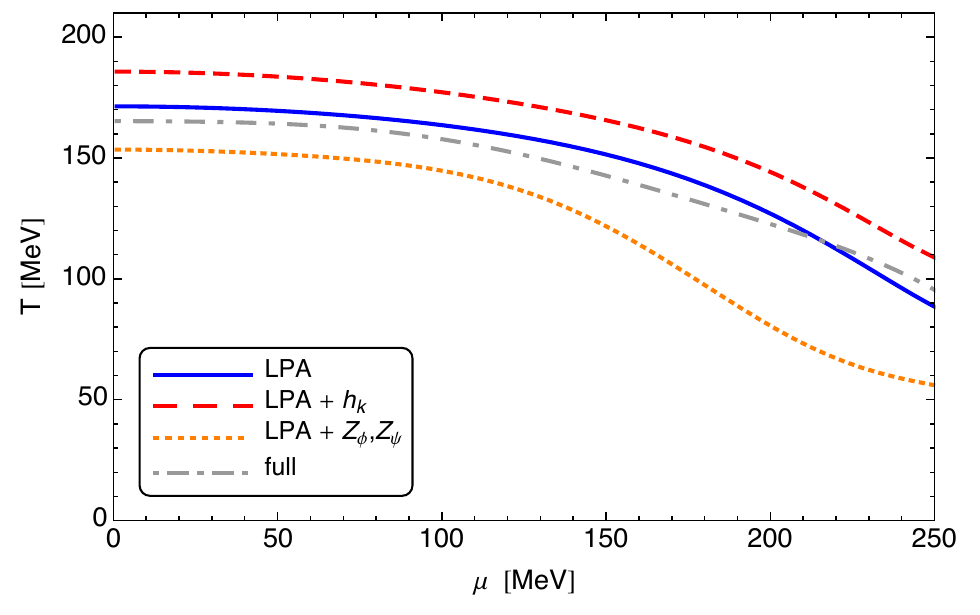}
  \caption{Comparison of the phase boundary for different truncations
    of the quark-meson model. The solid blue line corresponds to the
    local potential approximation (LPA), i.e. a truncation with only a
    running effective potential. For the dashed red curve also a
    running (field-independent) Yukawa coupling was taken into account
    and for the dotted orange curve we have running quark and meson
    wave function renormalisations and the effective potential, but a
    constant Yukawa coupling. The dot-dashed gray curve shows the full result of our model. Here, we defined the crossover transition via definition (i), see text.}\label{fig: cc}
\end{center}
\end{figure}
%

The inclusion of running wave function
renormalisations enhances the symmetry-preserving mesonic fluctuations
and therefore decrease the critical temperature. In turn, the
transition temperature is increased if a the running
Yukawa coupling is taken into account. This is shown in figure
\ref{fig: cc}. Note that we used the initial conditions specified in
\ref{subs: ini}. This ensures that for every truncation used in figure
\ref{fig: cc} we start with the same effective action at the initial
scale $\Lambda$. This approach is motivated by the fact that in
principle the initial conditions at $\Lambda$ are uniquely defined by
the solution of full QCD at scales $k\geq \Lambda$. 

\subsubsection{Curvature}
In order to determine the curvature of the chiral phase boundary, we
compute $T_c(\mu)/T_c(0)$ in a range $0\leq \mu/(\pi T_c(0))\lesssim
0.1$ and extract the curvature of the phase boundary at vanishing
chemical potential from these results. At small chemical potential the
phase boundary can be expanded in powers of $\mu^2$ as follows:
\begin{align}\label{eq: curvf}
  \frac{T_c(\mu)}{T_c(0)} = 1-\kappa_\mu \left(\frac{\mu}{\pi T_c(0)}
  \right)^2+\mathcal{O}\biggl(\!\left(\frac{\mu}{\pi
      T_c(0)}\right)^4\biggr).
\end{align}
The curvature $\kappa_\mu$ depends on the number of colours $\Nc$, the
number of quark flavours $\Nf$ and the current quark mass or the pion
mass respectively, see e.g. \cite{Braun:2008pi}. But since all those
parameters are fixed in the present work, we do not study the effect
of variations of them. For a crossover transition the
curvature depends on the definition of the phase boundary. For our
result in comparison with lattice results and other RG calculations
see table \ref{tab: curv}. We extracted the curvature from a fit of the phase boundary according
to \eq{eq: curvf} for $\mu \in [0,20]$ MeV. The errors result from
fits with polynomials of the order $\mu^2,\,\mu^4,\,\mu^6$.

Compared to the curvature found in \cite{Braun:2011iz}, the inclusion
of higher order mesonic scattering processes and dressed quark and
meson propagators does not change the curvature much. This is related
to the observation that running wave function renormalisations and the
running Yukawa coupling have opposing effects on the phase boundary,
see also \Fig{fig: cc}.

\begin{table}[t]
\begin{tabular}{|l| l| c| l|}
  \hline
  \quad \, method & boundary def. & mass & \,\;\quad$\kappa_\mu$\\
  \hline\hline
  Lattice: $i\mu$ \cite{deForcrand:2002ci}& plaquette susc. & 
  $am=0.025$ & $ 0.500(54) $\\
  \hline
  FRG: LPA \cite{Schaefer:2004en} & $\min_T\{\rho_0'(T)\}$ &
  chiral limit & $ 1.135 $\\
  \hline
  FRG: LPA \cite{Braun:2011iz}& $\min_T\{\rho_0'(T)\}$ & $m_\pi=
  138$ MeV & $ 1.375(63) $\\
  \hline
  this work: LPA & $\max_T\{\chi_{\bar{q}q}\}$ & $m_\pi=138$ MeV 
& $ 1.397(1) $\\
  \hline
  \multirow{3}{20mm}{this work: full model (\ref{eq:model})}  
& $\min_T\{\rho_0'(T)\}$  & 
  \multirow{3}{22mm}{$m_\pi=138$ MeV} & $ 1.397(2) $\\
  & $\max_T\{\chi_{\bar{q}q}\}$ & & $1.418(13)$ \\
  & $\min_T\{\bar V''(\bar\rho_0) \}$ & & $0.794(1)$ \\
  \hline
\end{tabular}
\caption{This table shows the curvature  of the chiral phase 
  boundary for $N_f=2$ quark flavours 
  obtained from various 
  methods. $am$ is the lattice spacing times the degenerate 
  current quark mass. The last three rows correspond to the different
  boundary definitions we employed in this work.}
\label{tab: curv}
\end{table}

Owing to our findings in the previous section we certainly need to use
the same definition of the phase boundary as in
\cite{deForcrand:2002ci} in order to do a sensible comparison with the
lattice results. But since they used the plaquette susceptibility for
the definition of the critical temperature, a direct comparison is
difficult since gluonic quantities are not directly accessible in our
model. We therefore displayed the results for the curvature for
different boundary definitions. We see that while the curvatures
extracted from the chiral order parameter and the chiral
susceptibility are very similar but much larger than the lattice
results, the curvature from the quartic meson coupling is close the
lattice result. We see that these results very much depend on the
specific definition of the crossover temperature, in line with our
findings in the previous section.

We note that it was observed for QM-model studies that the curvature
increases with increasing pion mass \cite{Braun:2011iz}, which
explains the difference between the curvature found in
\cite{Schaefer:2004en} and in \cite{Braun:2011iz}, where very similar
truncations were used but one in the chiral limit and the other at
realistic pion masses. This is in contrast to the general expectation
that the system gets less sensitive to the chemical potential for
larger current quark mass.


\section{Conclusions and Outlook}\label{sec: concl}

In this work, we have investigated the impact of higher order mesonic
scattering processes on the matter sector of two-flavour QCD at finite
temperature and quark chemical potential. Quantum, temperature and
density fluctuations have been taken into account within a
renormalisation group analysis of a quark-meson model. In particular,
we have introduced for the first time a meson-field dependent Yukawa
coupling. The effect of higher order meson-meson and quark-meson
operators has been systematically studied by expanding both the Yukawa
coupling and the effective potential in orders of the meson
fields. These higher order operators play a quantitatively important
role for the chiral phase transition. Furthermore, we observed that
these operators become increasingly irrelevant with increasing order
of the meson fields, see \Fig{fig: conv}. This indicates a rapid
convergence of the expansion scheme we used and allows us to have
certain control over the quantitative precision of our results.

We have computed the phase diagram of the chiral transition at finite
temperature and quark chemical potential, see \Fig{fig:
  pb}. Owing to the explicit $O(4)$-symmetry breaking which is
directly related to finite current quark masses we see a broad
crossover phase transition for $\mu < 291$ MeV. Crossover temperatures
cannot be defined uniquely. In the present work we have compared
standard definitions for the phase boundary and the corresponding
temperatures show the expected large deviations. In particular this
implies large differences in both the critical temperature at
vanishing chemical potential and the curvature. In the chiral limit,
all definitions provide the same results. 

At large chemical potential close to the critical point the phase
boundary is again uniquely defined since the crossover gets steeper in
this region. Even though we employed a local expansion of the
effective potential in this work, our particular expansion scheme
allowed us to resolve some global features of the effective
potential. This way we could capture the first order phase transition
for not too small temperatures and we found a critical endpoint at
$(T_\text{CEP},\mu_\text{CEP})=(50,291)\, \text{MeV}$.

Note, however, that at large chemical potential and small temperatures
quark-meson models in the present approximation are not expected to
give an accurate description of the QCD phase structure since diquark
and baryonic fluctuations should play an important role in this
region. Within the present approximation they are only taken into
account implicitly, the improvement of the present work in this
direction will be discussed elsewhere.

\vspace{0.1cm} {\it Acknowledgments -} We thank Jens Braun, Lisa Marie
Haas, Tina K.~Herbst, Naseemuddin Khan, Mario Mitter, Daisuke Sato,
Bernd-Jochen Schaefer, Nils Strodthoff, and Masatoshi Yamada for
discussions and collaboration on related subjects. JMP thanks the
Yukawa Institute for Theoretical Physics, Kyoto University, where this
work was completed during the YITP-T-13-05 on 'New Frontiers in
QCD'. This work is supported by Helmholtz Alliance HA216/EMMI and by
ERC-AdG-290623.

\begin{appendix}

\section{Expansion scheme} \label{app: exp}

In this Appendix technical details and convergence properties of the
present expansion scheme are discussed. Most expansion schemes in the
literature are either based on a discretisation of the
effective potential in field-space or a Taylor expansion about the
scale dependent minimum of the effective potential. The latter
approach is very efficient for low-order truncations with many
different interaction channels and has been proven to be very
successful at the description of critical phenomena (see
e.g.~\cite{Litim:2002cf}). The former gives a very detailed global
picture of the effective potential and is therefore well suited to
study first order phase transitions. Our expansion scheme may be seen
as a compromise between both approaches without being numerically
extensive. It is easily possible to include various directions in 
parameter space into the truncation while maintaining global 
information about the effective potential to a good accuracy.

 \subsection{Background dependence}

%
\begin{figure}[t]
\begin{center}
  \includegraphics[width=0.98\columnwidth]{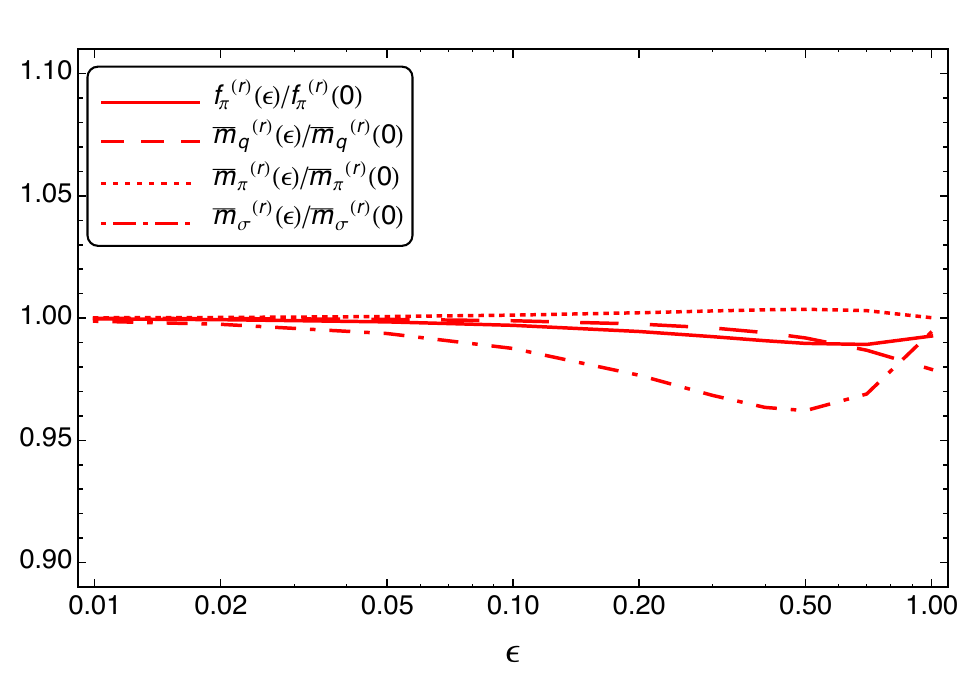}
  \caption{Dependence of the IR-observables on the offset parameter
    $\epsilon$. Our results for the observables are very robust with
    respect to variations of the expansion point if we take the
    corrections (\ref{eq: redef}) into account, implying a high degree of convergence of
    our expansion.}\label{fig: epsdep}
\end{center}
\end{figure}
%

Instead of doing an expansion about the scale-dependent minimum of the
effective potential $\rho_{0,k}$, we expand the non-renormalised
theory about a scale-independent field configuration $\kappa$, see
\eq{eq: zexp}, \eq{eq:vbar} and \eq{eq:barh}. Technically, the
advantage is that there is no unnecessary feedback from the expansion
point into the flow of higher order operators. In an expansion about
the minimum of the effective potential the flow of the minimum feeds
back into the flow of every higher order operator, see the discussion
below \eq{eq:lflow}. This feedback slows down numerical computations
and potentially leads to numerical instabilities. But even though the
minimum certainly is a distinct point in the effective potential, it
is by no means distinct in the flow of the effective potential. The
same holds true for the flow of the effective action in general. In
principle it is therefore irrelevant whether one solves the flow
equations with an expansion about $\bar\rho_{0,k}$ or any other point
in field space. $\bar\rho_{0,k}$ can always be extracted from
$\bar{V}_k(\bar\rho)$ from eq. (\ref{eq: min}) and enters the physical
parameters such as the physical masses,
$\bar{m}_k^{(\text{phys})}=\bar{m}_k(\bar\rho_{0,k})$.

There are two main restrictions we have for the choice of the
expansion point $\kappa$. The first and most important is that
$\kappa$ always has to be larger or equal to $\rho_{0,k}$ for small $k$. The reason
is that for $k\rightarrow0$ the effective potential becomes a convex
function of $\rho$ which is flat for $\rho<\rho_{0,k=0}$ and we can
not expect to capture the relevant features of the theory with an
expansion in a potentially flat region of the potential, especially
since all the physical information is stored in the effective
potential and its derivatives at the minimum. However, we can extract
all the information we need at much larger scales since the RG-flows
of the physical parameters stop at $k \approx m_{\pi}$. The remaining
flow for $k<m_\pi$ flattens the potential but leaves the physical
parameters unchanged.

Observables are extracted at the minimum of the potential at
$\bar\rho_{0,\text{IR}}(T,\mu)$. The present approximation has
field-independent wave function renormalisations. This introduces an
error which increases with the distance of the expansion point to the
minimum.  Consequently this leads to a finite radius of convergence in
$\rho$ about the minimum, leave aside general convergence issues of
the present Taylor expansion. Hence the expansion point should not
be too far away from the physically relevant region. This is assured
by choosing the expansion point close to the temperature and chemical
potential dependent IR-minimum:
\begin{align}
\bar\kappa_\text{IR}(T,\mu)=(1+\epsilon)\,\bar\rho_{0,\text{IR}}(T,\mu),
\end{align}
where $\epsilon$ is a small offset parameter.

The requirement of small $\epsilon$ is at least reduced qualitatively
if we would also take field-dependent wave function renormalisations
into account. In this work we only have considered wave function
renormalisation evaluated at the expansion point, see eq. (\ref{eq:
  zev}). Even though this is consistent with our expansion and the
proper way to define RG-invariant couplings which are also defined at
the expansion point, we expect some residual effects of the constant
wave function renormalisations on the physical quantities that are
defined at the minimum of the effective potential. In order to
partially compensate for this mismatch, we redefine the renormalised
IR-observables as follows:
\begin{align}\label{eq: redef}
\begin{split}
\bar{f}_\pi^{(r)}&=\sqrt{Z_{\phi,\text{IR}}(\bar\rho_{0,\text{IR}})/Z_{\phi,\text{IR}}(\bar\kappa_\text{IR})}\,\bar{f}_\pi,\\
 \bar{m}_{\phi}^{(r)}&=\sqrt{Z_{\phi,\text{IR}}(\bar\kappa_\text{IR})/Z_{\phi,\text{IR}}(\bar\rho_{0,\text{IR}})}\,\bar{m}_\phi,\\
 \bar{m}_{\psi}^{(r)}&=\left(Z_{\psi,\text{IR}}(\bar\kappa_\text{IR})/Z_{\psi,\text{IR}}(\bar\rho_{0,\text{IR}})\right)\,\bar{m}_\phi.
\end{split}
\end{align}
$Z_{\phi/\psi,\text{IR}}(\bar\rho_{0,\text{IR}})$ corresponds to the wave function
renormalisations at the IR minimum of the effective potential. It is
obtained from integrating the anomalous dimensions (\ref{etapsi}) and (\ref{etaphi}) 
at the physical point on the solution of the system at $\bar\kappa_k$. This 
ensures that the physical quantities are renormalised at the
physical point in the IR and furthermore allows us to examine the
robustness of our expansion even though we work with field-independent
wave function renormalisations.  For the sensitivity of our results on
$\epsilon$ with this correction, see \Fig{fig:
  epsdep}.  We see that the present expansion is surprisingly robust,
even though we dropped the field-dependence of the wave
function renormalisations. This observation is also reflected in \Fig{fig:PDEvsTaylor}. Furthermore, given the fact that we only made 
a simple adjustment to the wave function renormalisations in order to define 
the physical observables, the robustness of our expansion already implies only 
a mild dependence of the wave function renormalisations on the meson fields. 
In the expansion (\ref{eq: zexp}) the zeroth order term certainly depends on the 
expansion point but already the first order seems to give only a small correction, 
otherwise we would see a much stronger dependence on the expansion point in 
\Fig{fig: epsdep}.

\subsection{Consistency checks}

\begin{figure}[t]
\begin{center}
  \includegraphics[width=0.98\columnwidth]{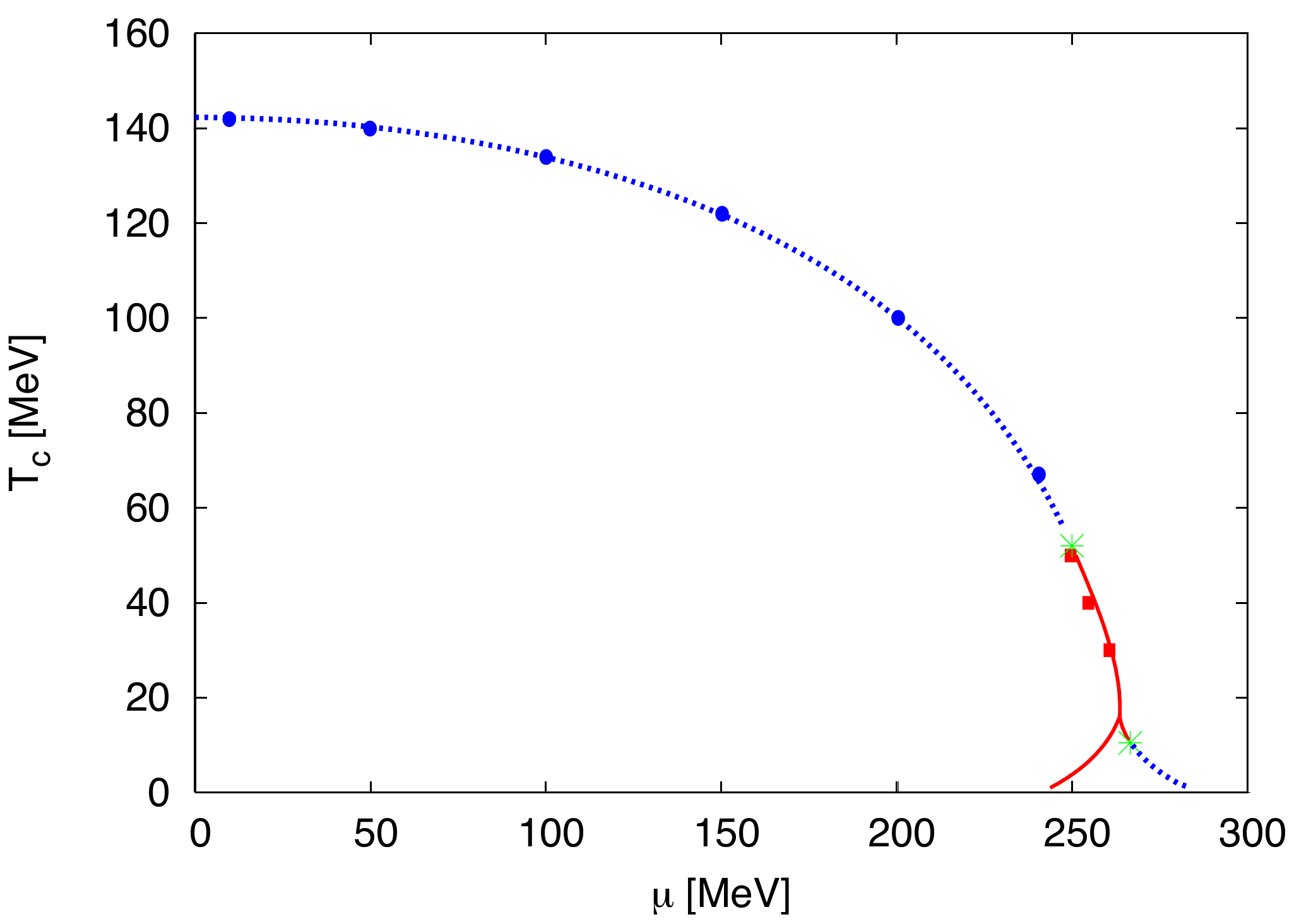}
  \caption{Phase diagram of the quark-meson model in the chiral limit
    from \cite{Schaefer:2004en}. The blue dots and red squares show
    the second and first order transition points we found using the
    identical model and initial conditions.}\label{fig: sw}
\end{center}
\end{figure}

Finally we present some checks concerning the validity of
our expansion. Very important in the context of this work is the
convergence of our expansion. This has already been demonstrated for
our model in section \ref{sec: conv}, see \Fig{fig: conv}. The
convergence of other observables may be faster or slower, but the
chiral condensate as a function of temperature and chemical potential
is certainly the crucial observable if one is interested in the chiral
phase transition.

We have determined the curvature for the quark-meson model as it is
used in \cite{Braun:2011iz} for infinite volume and found
\begin{align}
\kappa_\mu= 1.381(76),
\end{align}
which agrees with the result found in the reference.

Furthermore, we computed some points of the phase diagram of the
quark-meson model in the chiral limit with the truncation used in
\cite{Schaefer:2004en}:
\begin{align}\label{eq qmtrunc}
\begin{split}
\Gamma_k = &\int_0^\frac{1}{T}\!dx_0\int \!d^3x\Bigl\{\yb(\gamma_\mu 
\partial_\mu+\gamma_0\mu)\psi\Bigr.\\&+\Bigl.(\partial_\mu\phi)^2 
+V_k(\rho)+h 
\yb(\gamma_5\vec{\tau} \vec{\pi}+i\sigma\phi)\psi \Bigr\},
\end{split}
\end{align}
where only the effective potential is running (LPA). The result is
shown in \Fig{fig: sw}. At vanishing temperature and density the
convergence of the Taylor expansion in LPA has been checked in \cite{Papp:1999he}. 

One reads-off from \Fig{fig: sw} that the Taylor expansion reproduces
the full result for the second-order transition, the critical endpoint
and and the first part of the first-order transition to an accuracy of
about 1 MeV. If we go further along the first-order line, our result
starts to deviate from the result of \cite{Schaefer:2004en} and we are
not able to resolve the splitting of the phase diagram. In this
region, the distance between the global minimum of the effective
potential in the broken phase and the second minimum that emerges and
becomes the global minimum in the symmetric phase is fairly large and
seems to be larger than our radius of convergence. We note, however,
that we expanded the effective potential to order $N_V=7$ and that
higher orders in the expansion may resolve this problem.

In conclusion we see that our expansion scheme converges rapidly, is
insensitive to variations of the expansion point and is well
compatible with a grid solution of the effective potential for not too
small temperature and too large chemical potential, where we can not
compete with the resolution of the grid at the current stage. But
since quark-meson models do not have baryonic degrees of freedom, we
expect that these models are not valid models of QCD for small
temperature and large chemical potential anyway.

\section{Threshold Functions}\label{app:ThresholdFunc}

In the flow equations in section \ref{sec:frg} we used threshold
functions which contain the momentum integration, the summation over
the Matsubara modes and the regulator dependence of the propagators of
our model.

We use the following definitions for the meson and quark propagators:
\begin{align*}
\begin{split}
G_{\phi}(\bar{m}_{\phi,k}^2)&=\frac{1}{z_{\phi,k}\, \omega_n^2/k^2+x
\left(1+r_B(x)\right)+\bar{m}_{\phi,k}^2},\\
G_{\psi}(\bar{m}_{\psi,k}^2)&=\frac{1}{z_{\psi,k}^2 (\nu_n+i\mu)^2/k^2
+x\left(1+r_F(x)\right)^2+\bar{m}_{\psi,k}^2},
\end{split}
\end{align*}
where $x=\vec{q}^{\,2}/k^2$, $\omega_n=2 \pi n T$ is the bosonic
Matsubara frequency and $\nu_n=2\pi\left(n+\frac{1}{2}\right) T$ is
the fermionic Matsubara
frequency. $z_{\phi,k}=Z_{\phi,k}^\parallel/Z_{\phi,k}^\perp$ and
$z_{\psi,k}=Z_{\psi,k}^\parallel/Z_{\psi,k}^\perp$ give the ratios of
the wave function renormalisations parallel and perpendicular to the
heat bath. Within our approximations this ratio is one,
$z_{\phi,k}=z_{\psi,k}=1$.

We use the following regulators for mesons and quarks:
\begin{equation}\label{eq:regs}
\begin{split}
R_k^\phi=Z_{\phi,k}\, \vec{q}^{\,2}\,r_B(x),\\
R_k^\psi=Z_{\psi,k}\, \vec{\gamma}\vec{q}\,r_F(x).
\end{split}
\end{equation}
We use optimised regulator shape functions $r_{B/F}(x)$
\cite{Litim:2000ci} in this work:
\begin{align}\label{litR}
\begin{split}
r_B(x)=\left(\frac{1}{x}-1\right)\Theta(1-x),\\
r_F(x)=\left(\frac{1}{\sqrt{x}}-1\right)\Theta(1-x).
\end{split}
\end{align}
This choice of regulator shape functions allows us to evaluate
momentum integrals and Matsubara summation analytically.

The functions $l_0^{(B/F,d)}$ in $d$ space-time dimensions that appear
in equations (\ref{vflow}) and (\ref{hflow}) are related to
bosonic/fermionic loops and are defined as follows:
\begin{align*}
\begin{split}
&l_0^{(B,d)}(\bar{m}_{\phi,k}^2,\eta_{\phi,k};T)\\
&\quad=\frac{T}{2 k}\sum_{n\in \mathbb{Z}} \int\!dx x^{\frac{d-1}{2}}\left(
\partial_t r_B(x)-\eta_{\phi,k} r_B(x)\right) G_{\phi}(\bar{m}_{\phi,k}^2)\\
&\quad=\frac{2}{d-1}\frac{1}{\sqrt{z_{\phi,k}(1+\bar{m}_{\phi,k}^2)}}
\left(1-\frac{\eta_{\phi,k}}{d+1}\right)\\
&\qquad\times \left(\frac{1}{2}+n_B(T,\bar{m}_{\phi,k}^2)\right),
\end{split}
\end{align*}
and
\begin{align*}
\begin{split}
&l_0^{(F,d)}(\bar{m}_{\psi,k}^2,\eta_{\psi,k};T,\mu)\\
&\quad=\frac{T}{k}\sum_{n\in \mathbb{Z}} \int\!dx x^{\frac{d-1}{2}}
\left(\partial_t r_F(x)-\eta_{\psi,k} r_F(x)\right)\\
&\qquad\times (1+r_F(x))G_{\psi}(\bar{m}_{\psi,k}^2)\\
&\quad=\frac{1}{d-1}\frac{1}{\sqrt{z_{\psi,k}^2(1+
\bar{m}_{\psi,k}^2)}}\left(1-\frac{\eta_{\psi,k}}{d}\right)\\
&\qquad\times \left[1-n_F(T,\mu,\bar{m}_{\psi,k}^2)-
n_F(T,-\mu,\bar{m}_{\psi,k}^2)\right],
\end{split}
\end{align*}
where $n_B$ and $n_F$ are the Bose- and Fermi distribution respectively:
\begin{align*}
\begin{split}
n_B(T,\bar{m}_{\phi,k}^2) &= \frac{1}{\exp\left(\frac{k}{T} 
\sqrt{(1+\bar{m}_{\phi,k}^2)/z_{\phi,k}}\right)-1}\\
n_F(T,\mu,\bar{m}_{\psi,k}^2) &= \frac{1}{\exp\left(\frac{k}{T} 
\left(\sqrt{(1+\bar{m}_{\psi,k}^2)/z_{\psi,k}^2}-\frac{\mu}{k}\right)\right)+1}.
\end{split}
\end{align*}
The threshold functions $l_n^{(B/F,d)}$ which represent loops with
$(n+1)$ bosons/fermions are defined via:
\begin{align*}
\frac{\partial}{\partial m^2}l_n^{(B/F,d)}(m^2)=-(n+\delta_{n0})l_{n+1}^{(B/F,d)}(m^2).
\end{align*}
The threshold functions that appear in (\ref{hflow}) are related to
loops with fermion- as well as boson-propagators and are defined as
\begin{align*}
\begin{split}
&L_{(1,1)}^{(d)}\left(\bar{m}_{\psi,k}^2,\bar{m}_{\phi,k}^2,\eta_{\psi,k},\eta_{\phi,k};T,\mu\right)\\
&\quad=\frac{T}{2k}\sum_{n\in\mathbb{Z}}\int\! dx x^{\frac{d-1}{2}}
\Bigl[\left(\partial_t r_B(x)-\eta_{\phi,k} r_B(x)\right)\Bigr.\\
&\qquad\times G_{\phi}^2(\bar{m}_{\phi,k}^2)G_{\psi}(\bar{m}_{\psi,k}^2)+2(1+r_F(x))\Bigr.\\
&\qquad\times\Bigl.\left(\partial_t r_F(x)-\eta_{\psi,k} r_F(x)\right)
G_{\phi}(\bar{m}_{\phi,k}^2)G_{\psi}^2(\bar{m}_{\psi,k}^2)  \Bigr].
\end{split}
\end{align*}
By using the optimised regulator shape functions we can perform the
integration and summation analytically and find:
\begin{align*}
\begin{split}
&L_{(1,1)}^{(d)}\left(\bar{m}_{\psi,k}^2,\bar{m}_{\phi,k}^2,\eta_{\psi,k},\eta_{\phi,k};T,\mu\right)\\
&=\frac{2}{d-1} \left[\! \left(1-\frac{\eta_{\phi,k}}{d+1}\right)\!\mathcal{FB}_{(1,2)}+\!\left(1-\frac{\eta_{\psi,k}}{d}\right)\!\mathcal{FB}_{(2,1)} \right],
\end{split}
\end{align*}
where we defined the function
\begin{widetext}
\footnotesize
\begin{align}\label{eq: trF}
\begin{split}
&\mathcal{FB}_{(1,1)}\left(\bar{m}_{\psi,k}^2,\bar{m}_{\phi,k}^2;T,\mu\right) = \frac{T}{k}\operatorname{Re}\left[\sum_{n\in \mathbb{Z}}G_{\psi}(\bar{m}_{\psi,k}^2)G_{\phi}(\bar{m}_{\phi,k}^2)\right]\\
&\quad= \operatorname{Re}\left\{\frac{1}{2\sqrt{1+\bar{m}_{\phi,k}^2}}\left(n_B(T,\bar{m}_{\phi,k}^2)+\frac{1}{2}\right) \left[\frac{1}{\bar{m}_{\psi,k}^2+1-\left(\mu/k-i\pi
T/k-\sqrt{1+\bar{m}_{\phi,k}^2}\right)^2} + \frac{1}{\bar{m}_{\psi,k}^2+1-\left(\mu/k-i\pi T/k+\sqrt{1+\bar{m}_{\phi,k}^2}\right)^2}\right]\right.\\
&\qquad\left.-\frac{1}{2\sqrt{1+\bar{m}_{\psi,k}^2}}\left(n_F(T,\mu,\bar{m}_{\psi,k}^2)-\frac{1}{2}\right) \frac{1}{\bar{m}_{\phi,k}^2+1-\left(\mu/k-i\pi T/k-\sqrt{1+\bar{m}_{\psi,k}^2}\right)^2}\right.\\
&\qquad\left.-\frac{1}{2\sqrt{1+\bar{m}_{\psi,k}^2}}\left(n_F(T,-\mu,\bar{m}_{\psi,k}^2)-\frac{1}{2}\right) \times\frac{1}{\bar{m}_{\phi,k}^2+1-\left(\mu/k-i\pi T/k+\sqrt{1+\bar{m}_{\psi,k}^2}\right)^2}\right\}.
\end{split}
\end{align}
\end{widetext}
These mixed diagrams are responsible for the complex valued Yukawa
coupling and quark anomalous dimension, see section \ref{sec:frg}. It
is therefore sufficient to consider only the real part of this
contributions in order to render those functions real.

The functions $\mathcal{FB}_{(m,n)}$ which represent the Matsubara
summation of loops with $m$ fermion propagators and $n$ boson
propagators can be obtained from $\mathcal{FB}_{(1,1)}$ by
differentiation with respect to the masses:
\begin{align*}
\begin{split}
\frac{\partial}{\partial \bar{m}_{\psi,k}^2}\mathcal{FB}_{(m,n)}
&=-m\,\mathcal{FB}_{(m+1,n)}\\
\frac{\partial}{\partial \bar{m}_{\phi,k}^2}\mathcal{FB}_{(m,n)}
&=-n\,\mathcal{FB}_{(m,n+1)}.
\end{split}
\end{align*}

The function $\mathcal{BB}$ encodes the Matsubara summation of loops
with two different meson propagators are defined as:
\begin{align*}
\begin{split}
&\mathcal{BB}_{(1,1)}(\bar{m}_{\phi_1,k}^2,\bar{m}_{\phi_2,k}^2;T,\mu)\\
&\quad=\frac{T}{k}\sum_{n\in \mathbb{Z}}G_\phi(\bar{m}_{\phi_1,k}^2)G_\phi(\bar{m}_{\phi_2,k}^2)\\
&\quad = \frac{1}{(\bar{m}_{\phi_2,k}^2-\bar{m}_{\phi_1,k}^2)\sqrt{1+\bar{m}_{\phi_1,k}^2}}\left(n_B(\bar{m}_{\phi_1,k}^2)+\frac{1}{2}\right)\\
&\qquad+\frac{1}{(\bar{m}_{\phi_1,k}^2-\bar{m}_{\phi_2,k}^2)\sqrt{1+\bar{m}_{\phi_2,k}^2}}\left(n_B(\bar{m}_{\phi_2,k}^2)+\frac{1}{2}\right),
\end{split}
\end{align*}
and
\begin{align*}
\begin{split}
\frac{\partial}{\partial \bar{m}_{\phi_1,k}^2}\mathcal{BB}_{(m,n)}&=-m\,\mathcal{BB}_{(m+1,n)}\\
\frac{\partial}{\partial \bar{m}_{\phi_2,k}^2}\mathcal{BB}_{(m,n)}&=-n\,\mathcal{BB}_{(m,n+1)}.
\end{split}
\end{align*}

The Matsubara summation of loops with several identical fermions is encoded in:
\begin{align*}
\begin{split}
&\mathcal{F}_{(1)}(\bar{m}_{k,\psi}^2;T,\mu)=\frac{T}{k}\sum_{n \in \mathbb{Z}} G_\psi(\bar{m}_{k,\psi}^2)\\
&=\frac{1}{2\sqrt{1+\bar{m}_{\psi,k}^2}}\left[1-n_F(T,\mu,\bar{m}_{\psi,k}^2)-n_F(T,-\mu,\bar{m}_{\psi,k}^2)\right]
\end{split}
\end{align*}
and
\begin{align*}
\frac{\partial}{\partial \bar{m}_{\psi,k}^2}\mathcal{F}_{(n)}&=-n\,\mathcal{F}_{(n+1)}.
\end{align*}
Note that this function is implicitly contained in the threshold
function $l_{n}^{(F,d)}$ that appears in the flow of the effective
potential.

\section{Convexity for $\rho<\rho_0$}\label{app:con}

Here we present the detailed discussion of the results outlined in Section~\ref{subs:conlarge}.
The following is short of a full proof which is
beyond the scope of the present work. Here we are rather interested in
an explanation of the properties of the solution found in the present
work. Nonetheless the present analysis outlines the complete analysis
necessary for the full proof.

For finite $k$ there
is a region $\rho<\rho_s\leq \rho_0$ where all the curvature masses
$\bar m^2$ in \eq{eq:mmass} are negative,
\begin{equation}\label{eq:negcurv}
-1< \0{V_{k}'(\rho)}{k^2}<0\, \quad {\rm and} \quad -1 < 
\0{ V_{k}'(\rho)+2 \rho V''(\rho) }{ k^2 }<0\,, 
\end{equation}
for $\bar m^2_{k,\pi}$ and $\bar m^2_{k,\sigma}$ respectively.  Note
that the pion mass, $\bar m^2_{k,\pi}$, is already negative for
$\rho<\rho_0$. At the lower bound, $\bar m^2_{k,\sigma/\pi}=-1$, the
flow exhibits a singularity. However, due to the convexity-restoring
property of the flow arranges this bound is never saturated and
convexity is approached smoothly for $k\to0$, see 
\cite{Litim:2006nn}. This formal property has the practical
consequence that it i.e.\ implies for the flow of $m^2_{k,\pi}$
derived from (\ref{vflow}) that
\begin{eqnarray}\nonumber 
&&\hspace{-.5cm}   \lim_{k\to 0} \partial_t \bar m_{\pi,k}^2 
=\lim_{k\to 0} \partial_t \0{V_k'(\rho<\rho_0)}{k^2}\\\nonumber 
  &=& -\0{1}{4 \pi^2}\Biggl[3\partial_\rho
        m_{\pi,k}^2\,\,
     l_1^{(B,4)}(\bar{m}_{\pi,k}^2)+\partial_\rho m_{\sigma,k}^2\,\,
   l_1^{(B,4)}(\bar{m}_{\sigma,k}^2) \\
  & & -4N_c N_f \partial_\rho m_{\psi,k}^2\,\,
     l_1^{(F,4)}(\bar{m}_{\psi,k}^2) \Biggr]-2\,\bar m_{\pi,k}^2 = 0\,.
\label{vpflow}\end{eqnarray}
The subscript $l_1$ in the threshold functions indicates the
derivative w.r.t.\ the respective $\bar m^2$, see
Appendix~\ref{app:ThresholdFunc}.  Here and in the following we omit
the dependence on the anomalous dimensions, the temperature and the
chemical potential of the threshold functions for the sake of
legibility. Note that seemingly also $\lim_{k\to 0} \partial_t \bar
m^2 <0$ is allowed but then $\bar m^2$ eventually becomes positive
which signals the symmetric phase.

First we note that the fermionic contribution in the last line of
\eq{vpflow} vanishes in the limit $k\to 0$: For finite quark mass
function, $m_{\psi,k\to0}^2>0$, the threshold function vanishes,
$l_1^{(F,4)}\propto (m_\psi^2)^{-3/2}$, with cubic powers of $k$. In
turn, for vanishing quark mass function, $m_{\psi,k}^2\propto
k^\gamma\to 0$ for $k\to 0$, and $\partial_\rho m_{\psi,k\to0}^2 =0$
(no oscillation of $m_{\psi,k\to0}^2$ with period $\rho/k^\gamma$),
the threshold function stays finite, $l_1^{(F,4)}(m_\psi^2)
<l_1^{(F,4)}(0)= 1/3$. In either case the fermionic contribution
vanishes.

Hence, in the limit $k\to 0$ and for $\rho<\rho_0$ the flow of the
mesonic effective potential is dominated by the mesonic fluctuations
and reduces to that of an $O(4)$-model. Self-consistency of the
constraint \eq{vpflow}, the similar one for $\bar m^2_{k,\sigma}$, and
\eq{eq:negcurv} leads to
\begin{equation}\label{eq:proximity} 
  \lim_{k\to 0}\0{1}{1+\bar m^2_{\sigma/\pi,k} (\rho<\rho_s)}
  =\0{c_{\sigma/\pi}(\rho)}{k^{2+\alpha}}> 0\,, 
\end{equation}
with some constant $c_{\sigma/\pi}$ and $\alpha>0$, and 
\begin{equation}\label{eq:dmzero}
\partial_\rho m_{\sigma/\pi}^2(\rho<\rho_s)\propto k^{4+\alpha}\,, 
\end{equation} 
where we have assumed that the dominant sub-leading terms in $\bar
m^2$ carry a $\rho$-dependence. The threshold function $l_1^{(B,4)}$
scales with $(1+\bar m^2)^{-3/2}$ and hence we conclude that
\begin{equation}\label{eq:alpha} 
\alpha= 2\,, 
\end{equation} 
in line with the full analytic derivations in
\cite{LitimPawlowskiVergara}. \Eq{eq:negcurv} already induces a
scaling of $\partial_\rho m_{\sigma,\pi}^2(\rho<\rho_s)$ with at least
$k^2$ in the absence of oscillations in $\bar m^2$ with period
$\rho/k^2$. The lack of these oszillations can indeed be proven but
the details of this proof are beyond the scope of the present work
\footnote{Such an oscillation may be generated by an inadequate
  numerical implementation.}. The flow contributions in \eq{vpflow}
have to cancel the order $k^0$ contributions in $2 \bar m_\pi^2$. This
requires diverging threshold functions leading to \eq{eq:dmzero} which
implies $\bar m_{\sigma/\pi}^2= -1 +O(k^2)$. In turn this leads to the
same constant $c$ in \eq{eq:proximity} for $\sigma$ and $\vec\pi$
respectively. \Eq{eq:proximity} reflects the fact that the convexity
restoring property of the flow is driven by the denominators of the
threshold functions being close to the singularity.

For the behaviour of the fermionic two-point function
$\Gamma_{\psi,k}$ in the broken phase for $|\phi|\leq |\phi_0|$, we
resort to a more general argument. Its flow is dominated by the
diagrams with mesonic cutted lines: the lines with regulator
insertions are proportional to the mesonic propagators squared,
$G_{\phi,k}$, and hence diverge for $k\to 0$. Moreover, the fermionic
propagator obeys the flow equation
\begin{eqnarray}\nonumber 
\partial_t G_{\psi,k}[\Phi](p)& =& -\012 \Tr\, \left[ G_{k}\, \partial_t R_k\, G_k\,
\0{\delta^2 }{\delta\Phi^2} \right] G_{\psi,k}[\Phi](p)\\[2ex] 
& & -  \left( G_{\psi,k}\,\partial_t R^\psi_k\, G_{\psi,k}\right)[\Phi](p)\,,
\label{eq:fullflow} \end{eqnarray}
where $\Phi=(\psi,\bar\psi,\phi)$, see \cite{Pawlowski:2005xe}. For
momenta $p^2\gg k^2$, $|\phi|\leq |\phi_0|$, and $k\to 0$ this reduces
to 
\begin{equation}\label{eq:asymp} 
  \partial_t \0{1}{\Gamma_{\psi,k}^{(2)}[\phi](p)} = 
  -\012 \Tr\, G_{\phi,k}\, \partial_t R^\phi_{k}\, G_{\phi,k}\,\0{\delta^2 }{\delta\phi^2} 
\0{1}{\Gamma_{\psi,k}^{(2)}[\phi](p)}\,,
\end{equation}
where we have set $\psi=\bar \psi=0$, and $R_k(p^2\gg k^2)
\approx 0$.  The full fermionic two-point correlation function in the
background of constant mesonic fields $\phi$ reads
\begin{equation}\label{eq:ferm2}
  \Gamma_{\psi,k}^{(2)}[\phi](p) =  Z_\psi(\rho,p^2)\left( \not{\! p}  +
    i \bar h (\rho,p^2)
    \left[\sigma  - i \gamma_5 \vec \tau\vec \pi \right]\right)\,.
\end{equation}
at vanishing chemical potential, $\mu=0$. In \eq{eq:ferm2} we have
dropped the $k$-subscripts in $Z$ and $\bar h$ for the sake of
conciseness. Hence the full propagator in the background of constant
mesonic fields $\phi$ is expanded as
\begin{equation}\label{eq:fermproppara}
\0{1}{\Gamma_{\psi,k}^{(2)}[\phi](p)}  = A(\rho,p^2) \not{\! p} +   
B(\rho,p^2)\left(\sigma \id + i \gamma_5 \vec \tau\vec \pi\right)\,,
\end{equation} 
where the coefficient functions $A,B$ depend on both, $Z$ and $h$,
\begin{eqnarray}\nonumber 
  A(\rho, p^2)&=& \0{1}{Z_\psi(\rho,p^2) \left( p^2 + 2 
      \bar h (\rho,p^2)^2 \rho\right)}\,,
  \\[2ex]
  B(\rho,p^2)&=& A(\rho,p^2) \,\bar h (\rho,p^2) 
\label{eq:A+B}\end{eqnarray} 
Finally this leads to the differential equations 
\begin{subequations}
\begin{eqnarray} \label{eq:DiffA} 
  \partial_t A(\rho,p^2) &=& -\Bigl[N_\pi g_{\pi,k}(\rho)\,\partial_\rho\\ \nonumber 
&&  \hspace{.5cm} +
g_{\sigma,k}(\rho)\left(\partial_\rho+2 \rho \partial_\rho^2\right)\Bigr]
\,A(\rho,p^2)\,,\\[2ex]\label{eq:DiffB}
\partial_t B(\rho,p^2) &=& - \Bigl[N_\pi g_{\pi,k}(\rho)]\,\partial_\rho \\ 
&&\hspace{.5cm} +
g_{\sigma,k}(\rho)\left(3\partial_\rho +
2 \rho \partial_\rho^2\right)\Bigr]\, B(\rho,p^2)\,,
 \nonumber \end{eqnarray}
\label{eq:DiffAB}\end{subequations}
where $N_\pi$ is the number of pions, in the present $N_f=2$ case we
have $N_\pi=3$.  The $g_{\sigma/\pi,k}$ are the scalar parts of the
operator $G_{\phi,k}\, \partial_t R^\phi_{k}\, G_{\phi,k}$ projected
on the $\sigma$-meson and pion respectively.
\begin{equation}\label{eq:frho}
  g_{\sigma/\pi,k}(\rho)=\012 \left[G_k\, \partial_t R_k\,
G_k\right]_{\sigma\sigma/\pi\pi}(\rho)>0\,, 
\end{equation}
For $\rho<\rho_0$ $g_{\pi,k}$ diverges in 
the limit $k\to 0$, while $g_{\sigma,k}$ diverges for $\rho<\rho_s$, 
\begin{equation}\label{eq:fphi}
  g_{\pi,k}(\rho<\rho_0) \to \infty\,,\qquad
  g_{\sigma,k}(\rho<\rho_{s}) \to \infty\,, 
\end{equation}
Moreover, in the respective divergence regimes the $g_{\sigma/\pi,k}$ do not depend on the
  fermionic propagator in leading order. Hence is an
  external input for the differential equations \eq{eq:DiffAB}. It is
  here where the decoupling of the (leading part of the) flow equation
  for the effective potential from the fermionic diagrams comes handy.

  For a general class of $g_{\phi,k}$ the differential equations for
  $A(\rho,p^2)$, $B(\rho,p^2)$ have simple, attractive fixed point
  solutions for $k\to 0$ and $\rho<\rho_0$,
\begin{equation}
\partial_\rho A_{k=0}(\rho,p^2)= 0\,, \qquad \qquad 
\partial_\rho B_{k=0}(\rho,p^2)= 0\,. 
\end{equation} 
It is also easily seen that for non-trivial positive boundary
conditions the coefficient functions $A,B$ approach constants given by
their values at the minimum $\phi_0$ in terms of $Z_\psi(\phi_0,p^2)$ and 
$\bar h(\phi_0,p^2)$. This entails that 
\begin{equation}\label{eq:limbarh} 
 \bar h(\rho\leq \rho_0,p^2)= \bar h(\rho_0,p^2)
\end{equation}
and hence 
\begin{equation}\label{eq:Z} 
  Z_\psi(\rho\leq \rho_0,p^2)= Z_\psi(\rho_0,p^2)\0{p^2 +
    2  \bar h(\rho_0,p^2)^2\rho_0}{
    p^2 + 2 \bar h(\rho_0,p^2)^2\rho}\,.
\end{equation}
Note that the prefactor $Z_\psi(\rho_0,p^2)$, evaluated at $p=0$, is nothing but the 
wave function renormalisation used in the present work for the deduction of 
physical quantities. This full solution entails a mass gap for the quark propagator in the
broken phase: for non-vanishing momentum $p\neq 0$ the propagator trivially 
has no pole. For $p=0$ the wave function renormalisation is given by 
\begin{equation}\label{eq:Zp=0} 
 Z_\psi(\rho\leq \rho_0,0)= Z_\psi(\phi_0,0)\0{\rho_0}{\rho}\,.
\end{equation}
In \eq{eq:Zp=0} we have used that both, $\bar h(\rho_0,0)^2 > 0$ and
$Z_\psi(\rho\leq\rho_0,0)>0$, which follows from the analysis done in the
present paper. With \eq{eq:ferm2} this leads to 
\begin{equation}\label{eq:Gferm2p=0app}
  \Gamma_{\psi,k=0}^{(2)}[\phi](p=0)= i\, Z_\psi(\rho_0,0)   
\bar h (\rho_0,0)\rho_0\0{\sigma  - i \gamma_5 
\vec \tau\vec \pi }{\rho}\,.
\end{equation}
The norm of \eq{eq:Gferm2p=0app} is the $\rho$-dependent mass-gap of 
the propagator and is read-off from \eq{eq:Gferm2p=0app} as 
\begin{equation}\label{eq:ferm2p=0app}
  \bar m_\psi^2(\rho\leq \rho_0)= \0{ \|\Gamma_{\psi,k=0}^{(2)}[\phi](p=0)
\|^2}{ Z_\psi(\rho_0,0)^2}=
\sqrt{2 \rho_0}  \bar h (\rho_0,0) \0{\rho_0}{\rho}\,.
\end{equation}
We conlude that the field-dependent mass gap is minimised on the
equations of motion, $\rho=\rho_0$ and
\begin{equation}\label{eq:massgapapp} 
m_\psi^2(\rho\leq \rho_0) \geq m^2_{\psi,\rm gap}>0\,.  
\end{equation} 
Note also that the present scaling analysis is readily extended to
finite temperatures and densities. It also entails that the present
Tayor expansion in the mesonic field with fixed epxansion point nd at
$p=0$ is sufficient to extract the physics information. However, it
cannot in general reproduce the asymptotic behaviour for $k\to 0$ and
$\rho<\rho_0$ at one of the necessary condition for the full analysis,
$p\gg k$, does not hold. 

The above arguments can also be applied to the mesonic propagators for 
$k^2 \ll p^2 \ll m_{\sigma}^2$ with the parameterisation (at $\vec \pi=0)$ 
\begin{equation}\label{eq:mespropsp}
\CP_{\sigma/\pi}(\rho,p^2) = \0{1}{Z_\phi(\rho,p^2)\left(p^2+
m^2_{\sigma/\pi}(\rho)\right)}\,,
\end{equation} 
where $m^2_{\sigma/\pi,k}(\rho)$ does not depend on
momentum. Following the arguments used for deriving the flows
\eq{eq:DiffAB} for the coefficient functions of the fermionic
propagator we are led to the flow 
\begin{eqnarray}\label{eq:DiffC}
\partial_t \CP_{\sigma/\pi}(\rho,p^2) 
&=&  \\\nonumber 
& &\hspace{-2.5cm}-\Bigl[N_\pi g_{\pi,k}(\rho)\,\partial_\rho+
g_{\sigma,k}(\rho)\left(\partial_\rho+2 \rho \partial_\rho^2\right)\Bigr]
\CP_{\sigma/\pi}(\rho,p^2) \,,
\end{eqnarray}
For $\rho<\rho_s$ we have $m^2_{\sigma/\pi}<0$ (but $p^2 +
m^2_{\sigma/\pi}>0$) and both masses vanish in the limit $k\to 0$. We
therefore conclude that 
\begin{equation}\label{eq:mrhos}
m^2_{\sigma/\pi}(\rho<\rho_0)=0\,,\qquad 
Z_\phi(\rho<\rho_0,p^2)= Z_\phi(0,p^2)\,.
\end{equation}
At $\rho=\rho_0$ there is a discontinuity as $m^2_{\sigma}$ jumps 
to its physical value.

\end{appendix}
\bibliography{yukawa}

\end{document}